%% file: Main2.tex
\documentclass[nofootinbib]{article}

\usepackage{amssymb}

\usepackage{amsmath}

\usepackage{bm}

\usepackage{jheppub}

\allowdisplaybreaks

\include{mydefs}

\title{Cutting out the cosmological middle man: \\ {\Large General Relativity in the light-cone coordinates}}

\author{Ermis Mitsou$^1$,}
\emailAdd{ermitsou@physik.uzh.ch}
\author{Giuseppe Fanizza$^2$,} 
\emailAdd{gfanizza@fc.ul.pt}
\author{Nastassia Grimm$^1$,}
\emailAdd{ngrimm@physik.uzh.ch}
\author{Jaiyul Yoo$^{1,3}$}
\emailAdd{jyoo@physik.uzh.ch}

\affiliation{$^1$Center for Theoretical Astrophysics and Cosmology, Institute for Computational Science, University of Zurich, CH--8057 Z\"urich, Switzerland}
\affiliation{$^2$Instituto de Astrofis\'ica e Ci\^encias do Espaço, Faculdade de Ci\^encias da Universidade de Lisboa, Edificio C8, Campo Grande, P-1740-016, Lisbon, Portugal}
\affiliation{$^3$Physics Institute, University of Zurich, Winterthurerstrasse 190, CH-8057, Z\"urich, Switzerland}

\abstract{Analytical computations in relativistic cosmology can be split into two sets: time evolution relating the initial conditions to the observer's light-cone and light propagation to obtain observables. Cosmological perturbation theory in the FLRW coordinates constitutes an efficient tool for the former task, but the latter is dramatically simpler in light-cone-adapted coordinates that trivialize the light rays towards the observer world-line. Here we point out that time evolution and observable reconstruction can be combined into a single computation that relates directly initial conditions to observables. This is possible if one works uniquely in such light-cone coordinates, thus completely bypassing the FLRW ``middle-man'' coordinates. We first present in detail these light-cone coordinates, extending and generalizing the presently available material in the literature, and construct a particularly convenient subset for cosmological perturbation theory. We then express the Einstein and energy-momentum conservation equations in these coordinates at the fully non-linear level. This is achieved through a careful 2+1+1 decomposition which leads to relatively compact expressions and provides good control over the geometrical interpretation of the involved quantities. Finally, we consider cosmological perturbation theory to linear order, paying attention to the remaining gauge symmetries and consistently obtaining gauge-invariant equations. Moreover, we show that it is possible to implement statistical homogeneity on stochastic fluctuations, despite the fact that the coordinate system privileges the observer world-line.}

\begin{document}

\maketitle

\flushbottom

\section{Introduction}

The standard approach for analytical calculations in cosmology is perturbation theory around the homogeneous and isotropic solution. In the overwhelming majority of works in the literature, the space-time of that solution is described by the Friedmann-Lema\^itre-Robertson-Walker (FLRW) class of metrics
\beq \label{eq:FLRW}
\ed \bar{s}^2 = - N^2(t) \,\ed t^2 + a^2(t)\, \ed l^2 \, ,
\eeq
where $N$ is the lapse function, whose choice determines the time parametrization, $a$ is the scale factor and $\ed l^2$ is the line-element of a 3-dimensional maximally symmetric Euclidean space. The form \eqref{eq:FLRW} has the convenient property of making explicit the isometries of that space-time. Further choosing ``simple'' spatial coordinates for $\ed l^2$, the fluctuations on top of this idealized solution can be decomposed in the spectral and scalar-vector-tensor (SVT) fashion, thus revealing the independent degrees of freedom at the linear level. This approach is therefore optimal for solving perturbatively the equations of motion of a given theory. We will refer to this class of coordinate systems, i.e. those in which the metric is \eqref{eq:FLRW} plus small-amplitude fluctuations, as the ``FLRW coordinates''.

In relativistic cosmology, solving equations of motion is not sufficient for comparing theory and observation -- one must also solve further differential equations in order to obtain the observable quantities: e.g. redshift, luminosity distance, galaxy number density, etc. Unlike the equations of motion, which can be split into evolution equations and purely spatial (``constraint'') equations, the equations determining the cosmological observables are of light-like nature, as they control the propagation of photons along the past light-cone all the way up to the observer point. Therefore, in the FLRW coordinates the perturbative solutions for observables involve integrals along the past light-cone of the solutions of the equations of motion -- the higher the order in perturbations, the larger the number of nested integrals. The corresponding literature is abundant, with well-established results at the linear level and also an important amount of work on non-linear effects in the past decade.\footnote{See \cite{Mitsou:2019nhj} for a large selection of references. This effort in pushing analytical calculations to higher order in perturbation theory is motivated by the upcoming order-of-magnitude increase in the quantity and quality of observational data. Indeed, in order to correctly interpret this data, one requires theoretical predictions that match the precision set by the observational uncertainties.} 

The analytical expressions for the solutions of cosmological observables are fairly complicated already at linear order, essentially due to the plethora of terms corresponding to distinct relativistic effects, but nevertheless tractable in practice. However, they often become discouragingly complicated at non-linear orders in perturbations, making their numerical computation tedious and prone to mistakes, thus motivating the consideration of alternative approaches. Interestingly, it is one of the conceptual foundations of General Relativity (GR) -- the independence of physics on the choice of coordinate system -- that provides the most dramatic simplification. To be specific, consider a set of coordinates denoted by $\{ t, w, \te^a \}$, with $a \in \{ 1, 2 \}$, and the following line-element 
\beq \label{eq:GenLC}
\ed s^2 = - 2 N \Up \ed t \, \ed w + \Up^2 \ed w^2 + \ga_{ab} \( \ed \te^a - U^a \ed w \) \( \ed \te^b - U^b \ed w \) \, .
\eeq
We further require that $N, \Up > 0$, that $\ga_{ab}$ is a Euclidean metric and that the $t,w = {\rm constant}$ hypersurfaces have spherical topology, so that the $\te^a$ are angular coordinates. As we will show, any space-time can be brought to this form by an appropriate coordinate transformation, at least in some region of the full manifold. Equation \eqref{eq:GenLC} therefore corresponds to a coordinate choice, not a restriction to a particular class of space-times, and we will refer to it as the ``light-cone'' coordinates (LC). The particular case $N=1$ is already known as the ``geodesic light-cone'' coordinates (GLC)~\cite{Gasperini:2011us}. Clearly, no choice of functions $\{ N, \Up, U^a, \ga_{ab} \}$ allows one to obtain the FLRW form \eqref{eq:FLRW}, because of the presence of the off-diagonal term $\sim \ed t\, \ed w$ and the absence of the $\sim \ed t^2$ term. Therefore, this choice of coordinates is radically different from the ones that are usually considered in cosmology, in the sense that it cannot be obtained by adding small metric fluctuations to \eqref{eq:FLRW}. 

So how does \eqref{eq:GenLC} simplify the computation of observables? The answer is two-fold:

\begin{itemize}

\item
The past light-cone of the observer, on which all of the observable information lies, is simply given by a hypersurface of the form $w = {\rm constant}$.

\item
The angular coordinates of the incoming photons on such light-cones are constant all along their trajectory, i.e. the light-rays reaching the observer are straight lines in coordinate space, just as in flat space-time.

\end{itemize}
The first point implies that the LC system is adapted to the ``input-output'' aspect of cosmology: the input data consists of initial conditions at some $t = {\rm constant}$ hypersurface in the early universe (e.g. after inflation or recombination), while the output data consists of the observable information, now also lying at a single coordinate value hypersurface $w = {\rm constant}$. The second point is the one which simplifies the actual computation of the observables: their expressions no longer contain integrals along the line of sight, since these integrals were precisely taking into account the bending and deformation of incoming light-beams due to curvature in the FLRW coordinates \cite{BenDayan:2012pp,BenDayan:2012wi,Fanizza:2013doa,DiDio:2014lka}. Thus, in LC coordinates the observables depend solely on fields evaluated at the source and the observer points, meaning that the computational cost of solving some differential equations (multiple nested integrals) is dramatically reduced down to some algebraic operations and possibly a few derivatives. 

Having mentioned the advantages, let us now turn to the price one must pay in return. First, from the second bullet point above we infer that the LC coordinates can only cover a finite patch around the observer point, because they break down as soon as the first caustic of incoming light-rays occurs. Nevertheless, this is not a problem for perturbative computations, since the fluctuation fields are defined on the background space-time which can be covered globally. Second, at the level of the homogeneous and isotropic solution, these coordinates obscure the translational part of the isometries. This is essentially due to their unavoidable angular nature, since they represent light-cone hypersurfaces and therefore privilege a given point in space, at every given time $t$, the ``observer''. Without access to the full spectral/SVT decomposition that the simpler 3d coordinates provide, one can only achieve a partial decoupling of the different fluctuation components at the linear level. These equations will therefore be a bit harder to solve, but still solvable, as we will illustrate. Finally, when considering stochastic fluctuations, the fact that the translational isometries look complicated in spherical coordinates might make the implementation of statistical homogeneity less obvious a priori. As we will see, however, this apparent problem can be resolved relatively easily and the LC approach does not require sacrificing the cosmological principle.

Until now the use of LC coordinates has been restricted to the following ``hybrid'' approach: work with FLRW coordinates to compute time-evolution from the early universe to the observer's light-cone (e.g. FLRW-longitudinal gauge) and then switch to (G)LC through a coordinate transformation to compute the observables \cite{BenDayan:2013gc,Fanizza:2014baa,Marozzi:2014kua,DiDio:2014lka,DiDio:2015bua,Fanizza:2015gdn,Fanizza:2015swa,Marozzi:2016uob, Marozzi:2016qxl,Fleury:2016htl,Fanizza:2018qux}. The rationale behind this approach is that each coordinate system is the most suited for the task at hand: FLRW coordinates simplify the description of time evolution, while LC coordinates simplify observable reconstruction. However, in proceeding so one is essentially trading the complication of solving the observable differential equations in the FLRW coordinates for the complication of performing the coordinate transformation between standard and LC. As a result, the computational complexity is not drastically reduced, but is merely displaced to a different stage of the overall calculation. 

Our central observation here is that the FLRW coordinates appear as an unnecessary ``middle man'' in the approach described above. Indeed, the $t$ coordinate of the LC system \eqref{eq:GenLC} parametrizes space-like Cauchy surfaces, so one can very well describe time evolution directly within that system. The aim of this paper is to express the minimal set of equations of motion that one requires in relativistic cosmology in the LC coordinates. Writing down equations in a given coordinate system is a long but straightforward task. However, if one proceeds in a brute-force manner, then one quickly realizes that the complexity of the resulting expressions renders them very hard to handle, especially if one takes into account the fact that we are interested in higher-order perturbation theory. Instead, here we will employ standard tools of differential geometry to obtain relatively compact expressions involving mathematical objects with a clear geometrical interpretation and readily solvable at the linearized level. The procedure can be summarized in the following three steps: 
\begin{itemize}

\item
We perform a standard $3+1$ space/time decomposition of the manifold (also known as ``ADM'' decomposition \cite{Arnowitt:1962hi}), in which the $N$ field of \eqref{eq:GenLC} appears as the ADM lapse function.

\item
We perform a further $2+1$ angles/radius decomposition of the space-like hypersurfaces, thus leading to a $2+1+1$ angles/radius/time decomposition of the full manifold.

\item
We identify the particular choice of ADM shift vector $N^i$ which leads to the LC form \eqref{eq:GenLC}. The norm of $N^i$ is the limit case $N_i N^i = N^2$ in which $\pa_t$ becomes light-like, while the direction of $N^i$ is set to the radial direction of the $2+1$ decomposition, thus leading to light-cone hypersurfaces.

\end{itemize}
In particular, from the last point it is clear that \eqref{eq:GenLC} is a choice of coordinates, since the freedom of choosing the ADM shift is part of the freedom of choosing the coordinate system. 

The equations that we will decompose as described above are the Einstein and energy-momentum conservation equations. We will proceed in great detail, so that the same procedure can be repeated for other equations of motion that one encounters in cosmology, such as the ones of other fields (e.g. scalar and vector).\footnote{The case of Boltzmann distributions would require some extra structure, but would make use of the same geometrical construction.} The end-product of the paper is a set of evolution equations, for gravity and matter, and purely spatial ``constraint'' equations. We will derive them at the fully non-linear level, but we will also provide their expression to linear order in perturbation theory around the homogeneous and isotropic solution, leaving higher order results for future work. With these equations, given some initial data at an early $t = t_i$ hypersurface, one can evolve this information up to some future time after the observation event $x_o$ and then simply ``read-off'' the observables by collecting field values on the corresponding past light-cone $w = w_o$. One can understand these equations as the fusion of the differential equations describing time-evolution and observable reconstruction, so that solving them amounts to solving both problems simultaneously. Importantly, the linearized equations are not much harder to solve than their counterparts in the FLRW coordinates, as we show by explicitly solving part of them and providing a road-map for solving the rest. As a result, we expect a clear gain in efficiency. Finally, another original aspect of this paper will be to explore the consequences, and also highlight the practical advantages, of allowing $N \neq 1$. 

The paper is organized as follows. In section \ref{sec:extra} we provide the necessary information in order to properly define and handle the LC coordinates, proving in particular the claims made in the introduction. The original part of this section is the discussion of the $N \neq 1$ generalization of the already known GLC case, including its potential for more convenient gauge choices, and the generalization of the regularity conditions at the observer to arbitrary observer dynamics, i.e. including acceleration and rotation. In section \ref{sec:prel} we present the geometrical tools that we will use to derive the desired equations, i.e. $d+1$ decomposition and the Gauss-Codazzi-Mainardi equations. The reader who is familiar with this machinery can skip that section, although we recommend a quick look to get acquainted with our conventions and definitions. Next, in section \ref{sec:GRLC} we express the Einstein and energy-momentum conservation equations in LC form, that is, through a 2+1+1 angles/radius/time decomposition of the space-time manifold and a specific choice of the ADM shift vector. Then, in section \ref{sec:confvar} we perform a field redefinition that simplifies the cosmological perturbation theory, which we then consider in section \ref{sec:linear} to linear order. There we focus on the linear theory around the homogeneous and isotropic background and present a convenient decomposition of the fluctuations, paying particular attention to the remaining gauge symmetries and forming the corresponding gauge-invariant combinations. We then provide the corresponding linearized equations in terms of the gauge-invariant fields, thus obtaining a welcome consistency check, as well as a road-map for solving these equations in general. Finally, in section \ref{sec:statistics} we discuss the essential ingredients for working out the statistics of the corresponding stochastic fields, showing in particular that statistical homogeneity can indeed be implemented, and in section \ref{sec:conclusion} we conclude.

\section{The light-cone coordinates and the conformal Fermi subset}  \label{sec:extra}

\subsection{Basics}

Let us start by writing explicitly the inverse metric components of \eqref{eq:GenLC}
\beq
g^{tt} = - N^{-2} \, , \hspace{0.5cm} g^{tw} = -N^{-1} \Up^{-1} \, , \hspace{0.5cm} g^{ta} = - N^{-1} \Up^{-1} U^a \, , \hspace{0.5cm} g^{ww} = 0 \, , \hspace{0.5cm} g^{wa} = 0 \, , \hspace{0.5cm} g^{ab} = \ga^{ab} \, ,
\eeq
where $\ga^{ab}$ is the inverse of $\ga_{ab}$. We first note that the $t = {\rm const}.$ hypersurfaces are space-like, because their normal vector $\sim g^{\mu\nu} \pa_{\nu} t \equiv g^{\mu t}$ has negative norm $g_{\mu\nu} g^{\mu t} g^{\nu t} \equiv g^{tt} < 0$, so $t$ is a time coordinate. On the other hand, the $w = {\rm const}.$ hypersurfaces are light-like, because their normal vector $\sim g^{\mu\nu} \pa_{\nu} w \equiv g^{\mu w}$ has vanishing norm $g_{\mu\nu} g^{\mu w} g^{\nu w} \equiv g^{ww} = 0$. Taking into account the requirement that the $t,w = {\rm const.}$ hypersurfaces have spherical topology, we have that the $w = {\rm const}.$ hypersurfaces are light-cones and are therefore attached to some central world-line in space-time, henceforth simply referred to as the ``observer''. One must choose between future and past light-cones, otherwise we over-parametrize space-time, and in this context the obvious choice is past. The observer world-line is described by a relation between the two coordinates $\{ t, w \}$
\beq \label{eq:obsWL}
w = w_o(t) \, , 
\eeq
since the angular coordinates are not defined there, by definition. Also, the function $w_o(t)$ must be monotonic for each past light-cone to be uniquely labeled. In general we will use a ``$o$'' subscript to denote evaluation at the observer world-line. For a visual illustration of the space-time slicing described here, we recommend fig. 1 of \cite{Fleury:2016htl}.

We next define the following future-oriented normal vector field to the $w = {\rm const.}$ hypersurfaces
\beq \label{eq:kmu}
k^{\mu} := - g^{\mu\nu} \pa_{\nu} w \equiv - g^{\mu w} = N^{-1} \Up^{-1} \( 1, 0, \vec{0} \)^{\mu} \, ,  \hspace{1cm} g \( k, k \) \equiv 0 \, ,
\eeq
and note that, for light-like vectors, being ``normal'' to a surface in the above sense is synonym to being tangent to it. Further observing that the $k^{\mu}$ vector field is geodesic
\beq
k^{\nu} \na_{\nu} k^{\mu} \equiv 0 \, ,
\eeq
we conclude that it can be interpreted as the 4-momentum vector of the incoming photons to the observer, up to a constant factor. One can then also verify that
\beq
k^{\mu} \pa_{\mu} \te^a = g^{wa} = 0 \, ,
\eeq
meaning that the angular coordinates of these incoming photons are constant all along their trajectory. Therefore, the LC coordinates trivialize the light-propagation along the $w = {\rm const.}$ past light-cones. 

Let us now observe that the line-element \eqref{eq:GenLC} does not correspond to a complete gauge fixing, i.e. one can still perform some coordinate transformations without altering its form. These are given by the full time-reparametrization freedom
\beq \label{eq:trep}
t \to t'(t,w,\te) \, ,
\eeq
and the light-cone and light-cone-dependent angle reparametrizations
\beq \label{eq:RGT}
w \to w'(w) \, , \hspace{1cm} \te^a \to \te'^a(w,\te) \, ,
\eeq
respectively. The former \eqref{eq:trep} is due to the fact that $N$ is the lapse function of the ADM decomposition, which is therefore completely free to choose. In the GLC case, where $N = 1$, the freedom \eqref{eq:trep} is reduced down to $t \to t + {\rm constant}$. The other two transformations \eqref{eq:RGT} do not involve a full space-time function and are therefore ``residual'' gauge transformations.\footnote{See \cite{Fleury:2016htl} for a detailed discussion of these freedoms in the GLC context.} 

Let us also mention another closely related class of coordinate systems which predates GLC, the so-called ``observational coordinates'' \cite{ObsCoord} (see also \cite{Nugier:2013tca,Mitsou:2019nhj}), where the time coordinate $t$ is traded for a space-like coordinate. Thus, instead of slicing space-time into light-cones and spatial hypersurfaces, one slices it into light-cones and time-like cylinders. Here, however, we are interested in describing evolution in time of $t = {\rm constant}$ data, so the appropriate system is LC.

\subsection{The interpretation of $N$}

The freedom of choosing $N$ is related to the different ways of slicing space-time into $t = {\rm constant}$ hypersurfaces. To see this, we define their future-oriented unit-normed normal vector
\beq \label{eq:nmu}
n^{\mu} := -\frac{g^{\mu\nu} \pa_{\nu} t}{\sqrt{-g^{tt}}} = \( N^{-1}, \Up^{-1}, \Up^{-1} U^a \)^{\mu} \, ,  \hspace{1cm} g \( n, n \) \equiv -1 \, ,
\eeq
and note that its evolution in time is entirely controlled by $N$
\beq \label{eq:acc}
n^{\nu} \na_{\nu} n^{\mu} \equiv N^{-1} \( g^{\mu\nu} + n^{\mu} n^{\nu} \) \pa_{\nu} N \, .
\eeq
Physically, the vector field $n^{\mu}$ can be thought of as the 4-velocity field of a family of test observers. Consistently with the fact that motion at constant $w$ means light-like motion, we see that the future time-like motion \eqref{eq:nmu} requires a strictly positive $w$ component $n^w > 0$. Moreover, that equation also allows us to interpret $U^a$ as the angular velocity of said observer family. The vector field in \eqref{eq:acc} is then the family's 4-acceleration, meaning that $N$ gives us access to a broad set of possible dynamics. In particular, GLC corresponds to the case where the family is in free-fall $N = {\rm constant}$, hence the ``geodesic'' part of the name. The value $N = 1$ is then obtained through a constant rescaling of $t$, thus making that variable the proper time of the family. 

The member of that family that sits at the tip of the light-cones is nothing but {\it the} observer defined through \eqref{eq:obsWL}, i.e. the actual observer involved in the cosmological observations. In the works employing the GLC system so far, one usually considers the rest of the members of that family to play the role of the sources involved in cosmological observations. As a result, in GLC both the observer and the sources (associated with this coordinate system through $n^{\mu}$) are in free-fall, while in the more general LC case their dynamics is controlled by $N$. Here we will choose not to perform this identification for the sources, so that $N$ will be free to choose either for simplifying some particular computations or for simplifying some particular observable.\footnote{For instance, one can choose $N$ such that the corresponding $t$ variable coincides with a monotonic observable such as redshift or cosmological distances, or at least such that one of these quantities has zero fluctuations in cosmological perturbation theory.\label{ft:otherN}} This generalized description of sources does not spoil at all the advantages of the LC coordinates, i.e. the observables are still local functions of the fields. For instance, since the $k^{\mu}$ field \eqref{eq:kmu} denotes the 4-momentum of incoming photons on the light-cone, the redshift of a given source with 4-velocity $V_s^{\mu}(t,w,\te)$ observed by an observer sitting at the central point with 4-velocity $n^{\mu}_o(w)$ is given by the integral-free expression (in LC coordinates)
\beq \label{eq:redshiftgen}
1 + z_V(t,w,\te) := \frac{(k_{\mu} V_s^{\mu})(t,w,\te)}{(k_{\mu} n_o^{\mu})(w)} \equiv \Up_o(w) \, V_s^w(t,w,\te) \, .
\eeq 
In the special case where the $V^{\mu}$ values are taken out of the $n^{\mu}$ field \eqref{eq:nmu}, one recovers the well-known result from the GLC literature
\beq \label{eq:redshift}
1 + z_n(t,w,\te) = \frac{\Up_o(w)}{\Up(t,w,\te)} \, ,
\eeq  
which therefore holds for generic $N$. The already known expressions for the GLC angular diameter and luminosity distances \cite{BenDayan:2012pp,Fleury:2016htl}, Jacobi map \cite{Fanizza:2013doa} and galaxy number density \cite{DiDio:2014lka} will be generalized to arbitrary $V^{\mu}$ and $N$ in future work.

\subsection{The temporal gauge and rigid conformal time parametrization} \label{sec:tempgauge}

Let us next look at the light-cone reparametrization freedom in \eqref{eq:RGT}, which amounts to choosing the function introduced in \eqref{eq:obsWL} that describes the location of the observer world-line. We will consider for definiteness the simplest choice for cosmology that is the ``temporal gauge'' \cite{Fleury:2016htl}
\beq
w_o(t) = t \, ,
\eeq
so that the observer world-line is located at $w = t$. Let us now see how this translates as a condition on the metric components. We first note that, under a light-cone reparametrization $w \to w'(w)$ we have in particular
\beq \label{eq:Upwtrans}
\Up^{-1} \to \frac{\ed w'}{\ed w}\, \Up^{-1} \, , 
\eeq
so that the gauge can be fixed through a condition on $\Up$. More precisely, at the observer point the function $\Up(t,w,\te)$ has no angular dependence (see below) and reduces to a function of $t$ alone $\Up_o(t) := \Up(t,w_o(t))$. Alternatively, if we use the inverse function $w_o^{-1}$, then we can define a function of $w$ only $\Up_o(w) := \Up(w_o^{-1}(w),w)$, which is exactly the amount of information we can manipulate with $w \to w'(w)$ given \eqref{eq:Upwtrans}. We are therefore looking for a condition on $\Up_o$. Denoting by $x_o^{\mu}(\ta)$ the observer world-line, where $\ta$ is their proper time, the fact that $n_o^{\mu}$ is the observer 4-velocity 
\beq
\frac{\ed x_o^{\mu}}{\ed \ta}(\ta) = n^{\mu}(x_o(\ta)) \, ,
\eeq
along with the temporal gauge
\beq
x_o^t(\ta) = x_o^w(\ta) \, ,
\eeq
implies that 
\beq \label{eq:tempgauge}
\Up_o(t) = N_o(t) \, ,
\eeq
which is therefore the expression of the temporal gauge in terms of the metric components. In particular, note that the temporal gauge is not consistent with the GLC gauge $N = 1$ if $\Up_o \neq 1$. This is actually the case for the global description of the cosmological homogeneous and isotropic solution where $\Up$ is the scale factor $a(t)$. The temporal gauge then implies that we will be working with the conformal time parametrization in that context. 

Finally, note that \eqref{eq:tempgauge} is a condition on $\Up_o$, but $N$ is still free to choose thanks to the time reparametrization symmetry \eqref{eq:trep}. In particular, $N_o(t)$ is controlled by the transformations of the form $t \to f(t)$, and one must simply be careful to combine these with $w \to f(w)$ in order to preserve the temporal gauge. This freedom can be fixed within cosmological perturbation theory by setting 
\beq \label{eq:rigidgauge}
N_o(t) = a(t) \, ,
\eeq
to all orders in the perturbations, i.e. setting to zero the scale factor fluctuations at the observer, which is why we will name this the ``rigid conformal time'' parametrization. Thus, from now we will use the notation $a(t) := N_o(t) = \Up_o(t)$, i.e. even when working at the fully non-linear level, with this quantity then reducing to the scale factor when doing cosmological perturbation theory. In practice, this simply means that when we will perturb $N = a + \de N$ around the cosmological background, we will impose $\de N_o = 0$, and so on for $\Up$.

\subsection{Regularity conditions close to the observer}

With the temporal gauge the quantity 
\beq
r := w - t \, , 
\eeq
controls the spatial distance to the observer world-line if $t$ is kept constant, i.e. it is a ``radius'' coordinate. Assuming a regular space-time at the observer (e.g. no black hole), we can therefore express the metric components through a series of the form
\beq \label{eq:gexp}
g_{\mu\nu}(t,w,\te) = \sum_{n=0}^{\infty} g^{(n)}_{\mu\nu}(t,\te) \,r^n \, . 
\eeq
The reason for considering such an expansion is that the spherical nature of the LC coordinates constrains the first few coefficients $g^{(n)}_{\mu\nu}(t,\te)$. This is essentially due to the consistency requirement that no angular direction can be privileged at the origin $w = t$, assuming that space-time is regular there. A first regularity condition is that a tensor of the form $T_{a_1\dots a_n}$ starts at order $r^n$, because otherwise the invariant quantity $T_{a_1\dots a_n} \ed \te^{a_1} \dots \ed \te^{a_n}$ would not be well-defined at the observer. Here this implies
\beq \label{eq:U0ga0ga10}
U_a^{(0)} \equiv 0 \, , \hspace{1cm} \ga_{ab}^{(0)} \equiv \ga_{ab}^{(1)} \equiv 0 \, ,
\eeq  
where the angular indices are displaced using $\ga_{ab}$, so $U_a := \ga_{ab} U^b$. But there are also further regularity conditions on the first few non-trivial terms of each series. Some of them are intuitive, e.g. for angular scalars the zeroth order term cannot depend on angles, since it survives in the limit of zero distance to the observer, so here
\beq \label{eq:consconds0}
N^{(0)}(t,\te) = \Up^{(0)}(t,\te) = a(t) \, ,
\eeq
as already discussed in the previous subsection. However, most of the regularity conditions are not so transparent and must therefore be derived, as we do in detail in appendix \ref{app:obscoord}, using a generalization of the method employed in \cite{Fanizza:2018tzp}. To express these conditions we need to define a few objects: a time-dependent parametrization of the unit-sphere in Euclidean space $\hat{X}(t,\te) := \{ \hat{X}^i(t,\te) \}_{i=1,2,3}$, i.e. 
\beq \label{eq:XX1}
\hat{X} \cdot \hat{X} \equiv 1 \, , 
\eeq
and the corresponding induced metric and volume form on the unit-sphere
\beq \label{eq:qtiqdef}
q_{ab} := \pa_a \hat{X} \cdot \pa_b \hat{X} \, ,  \hspace{1cm} \ti{q}_{ab} := \sqrt{q}\, \vep_{ab} \, ,
\eeq
respectively. The simplest case is 
\beq \label{eq:standangparam}
\hat{X} = \( \sin \te \cos \vph, \sin \te \sin \vph, \cos \te \) \, , \hspace{1cm} q_{ab} = {\rm diag}(1, \sin^2 \te)_{ab} \, , \hspace{1cm} \ti{q}_{ab} = \sin \te\, \vep_{ab} \, ,
\eeq
but we will not commit to a particular $\hat{X}(t,\te)$ map yet. In terms of these quantities, the remaining regularity conditions read\footnote{The condition \eqref{eq:conscond1nt} invalidates the argumentation and conclusion of \cite{Mitsou:2017ynv}, which is based on the erroneous assumption that $\ga^{(2)}_{ab}$ can have a non-constant curvature, due to neglecting the regularity conditions at the observer.} 
\beq \label{eq:conscond1nt}
U_a^{(1)} = 0 \, ,  \hspace{1cm} \ga_{ab}^{(2)} = a^2 q_{ab} \, , 
\eeq
for the first non-trivial order and
\bea 
N^{(1)} & = & a^2 \vec{\al} \cdot \hat{X} \, , \label{eq:conscondN1} \\
\Up^{(1)} & = & a^2 \[ 2 Z + \vec{\al} \cdot \hat{X} \] \, , \label{eq:conscondUp1} \\
U_a^{(2)} & = & -\, a^3 \[ \pa_a Z + \ti{q}_a^{\,\,\,b} \pa_b \( W + \vec{\om} \cdot \hat{X} \) \] \, , \label{eq:conscondUa2} \\
\ga_{ab}^{(3)} & = & 2 a^3 \[ \( Z + \vec{\al} \cdot \hat{X} \) q_{ab} + \ti{q}_{(a}^{\,\,\,\,c} D_{b)} \pa_c W \] \, , \label{eq:conscondgaap3}
\eea
for the second one. Here $\vec{\al}(t)$ and $\vec{\om}(t)$ are the acceleration and angular velocity vectors of the observer in their proper reference frame, respectively, $W(t,\te) \sim \ti{q}^{ab} \pa_a \hat{X} \cdot \pa_b \pa_t \hat{X}$ fully captures time-dependence of $\hat{X}(t,\te)$, $Z(t,\te)$ is also an undetermined function and $D_a$ is the covariant derivative associated to $q_{ab}$. Evaluating \eqref{eq:acc} at the observer position one obtains $\de_w^{\mu}\, \vec{\al} \cdot \hat{X}$, which is consistent with the interpretation of this vector as the 4-acceleration of the observer. Equations \eqref{eq:consconds0}, \eqref{eq:conscond1nt} and \eqref{eq:conscondN1} to \eqref{eq:conscondgaap3} are the complete set of regularity conditions, i.e. there are no more constraints to higher orders.

\subsection{The non-rotational observational gauge}

We now consider the angular reparametrization freedom in \eqref{eq:RGT} and note that it must be fixed on practical grounds. Out of all the possible parametrizations of the sky, only a small subset corresponds to the angles that the actual observer uses in practice, since these angles are defined only up to a global rotation of the sky, not the full 2d diffeomorphism group. In particular, correlation functions and spectral analysis on the sphere implicitly make use of these ``observed'' angular coordinates. Moreover, these coordinates are usually defined with respect to distant (``at infinity'') reference objects in the sky, thus rotating in time so as to compensate the spinning of the observer $\vec{\om}(t)$. This extra requirement then determines the angular coordinates up to a time-independent rotation in the sky. The result is called the ``non-rotating observational gauge'' \cite{Fleury:2016htl} and has been constructed explicitly in \cite{Fanizza:2018tzp} for the $N = 1$ and $\vec{\om} = 0$ case. 

Once the regularity conditions \eqref{eq:conscond1nt} are given, it is easy to see how that gauge can be imposed. Following the same logic we use to fix $\Up_o(t)$ through $w \to w'(w)$, here too the freedom $\te^a \to \te'^a(w,\te)$ is equivalent to $\te^a \to \te'^a(t,\te)$, since $w = t$ at the observer, meaning that we can choose the map $\hat{X}(t,\te)$ freely. Since $\ga_{ab}^{(2)} = a^2(t)\, q_{ab}(t,\te)$ is the angular geometry infinitesimally close to the observer, it constitutes their ``sky'', so we must choose a $\hat{X}(t,\te)$ that reproduces the standard 2-metric
\beq \label{eq:std2metric}
q_{ab} = {\rm diag}(1, \sin^2 \te)_{ab} \, .
\eeq
This fixes the function $\hat{X}(t,\te)$ up to time-dependent rotations, which is exactly the freedom we need to compensate the observer's rotation. Setting without loss of generality $\vec{\om}(t) = \om(t) \( 0,0,1 \)$, we can consider the rotating generalization of \eqref{eq:standangparam}
\beq  \label{eq:rotX}
\hat{X}(t,\te) = \( \sin \te \cos \( \vph - \int^t \ed t'  a(t')\, \om(t') \), \sin \te \sin \( \vph - \int^t \ed t' a(t')\, \om(t') \), \cos \te \) \, , 
\eeq
which leads to the same 2-metric \eqref{eq:std2metric}, but also (see appendix \ref{app:obscoord})
\beq
W = - \, \vec{\om} \cdot \hat{X} \, ,
\eeq
which compensates indeed the effect of the observer's rotation in the metric components \eqref{eq:conscondUa2}  
\beq \label{eq:conscondUa2noom}
U_a^{(2)} = - a^3 \pa_a Z \, ,
\eeq
and also simplifies \eqref{eq:conscondgaap3} down to
\beq
\ga_{ab}^{(3)} = 2 a^3 \[ Z + \vec{\al} \cdot \hat{X} \] q_{ab} \, .
\eeq
Finally, in the less usual case where one defines the angles with respect to earth-fixed references, the choice is \eqref{eq:standangparam}, so that $W = 0$ and $U_a^{(2)} = - a^3 \[ \pa_a Z + \vec{\om} \cdot \ti{q}_a^{\,\,\,b} \pa_b \hat{X} \]$.

\subsection{One final simplification}

The gauges imposed so far are already known in the GLC context. As we will now show, however, the consideration of $N \neq 1$ leads to a very convenient and natural gauge choice, which is not reachable when $N = 1$. To see this, let us consider the coordinate transformation performed in appendix \ref{app:obscoord}, which relates the generalized Fermi normal coordinates $\{ T, \vec{X} \}$ to the LC ones $\{ t,w,\te \}$. After imposing the regularity conditions, the temporal part of the transformation \eqref{eq:Texp} reduces to
\beq
T(t,w,\te) = \int^t a(t') \, \ed t' + \[ \frac{1}{2}\, \dot{a}(t) - a^2(t)\, Z(t,\te) \] r^2 + \Ord(r^3) \, ,
\eeq
where we have used \eqref{eq:ZofT2}. The fact that the zeroth and first orders are completely determined is due to the fact that we have fixed $N^{(0)} = a$ and that $N^{(1)}$ is determined by the observer's acceleration \eqref{eq:conscondN1}. We then see that the second-order coefficient is essentially $Z$. Since we have only considered the first two non-trivial orders in the computation of appendix \ref{app:obscoord}, we do not know a priori whether $Z$ is free to choose or determined. Going to the next order, one can verify that $Z$ is indeed free to choose, because it must satisfy an equation which involves, not surprisingly, the free function $N^{(2)}(t,\te)$. Thus, $Z$ parametrizes the freedom of performing time-reparametrizations at that order, while remaining within the LC class of coordinates. Put differently, $Z$ can be brought to any desired function through a time-reparametrization of order $r^2$, and doing so simply amounts to fixing $N^{(2)}$. Given the regularity conditions \eqref{eq:conscondN1} to \eqref{eq:conscondgaap3}, a natural choice is then
\beq \label{eq:Fermi}
Z = 0 \, .
\eeq 
It is now clear why this is not possible to enforce in the GLC case, where $N^{(2)}$ is fixed to zero. Indeed, in \cite{Fanizza:2018tzp} the authors showed that $Z$, or more precisely their analogue of $\Up^{(1)}$, is fully determined by some components of the Riemann tensor at the observer position and in the Fermi coordinates. In contrast, here an analogous relation arises, but also involving the free function $N^{(2)}$. Setting \eqref{eq:Fermi} therefore determines $N^{(2)}$ in terms of the Riemann tensor components in the Fermi coordinates, i.e. this is not a regularity condition, but just means that $N^{(2)}$ is no longer free to choose.

\subsection{Full gauge recap: the light-cone conformal Fermi coordinates} \label{sec:LCCF}

Starting with the generic LC coordinates \eqref{eq:GenLC}, we have further specified these coordinates through successive conditions, either for computational convenience, or out of physical requirements. The first kind of conditions are the temporal gauge \eqref{eq:tempgauge}, rigid conformal time parametrization \eqref{eq:rigidgauge} and also \eqref{eq:Fermi}, while the second kind is the non-rotating observational gauge \eqref{eq:std2metric} and \eqref{eq:rotX}. All of these conditions can be collectively expressed through the form of the lowest-order coefficients of the metric 
\bea 
N & = & a \[ 1 + a r \vec{\al} \cdot \hat{X} + \Ord(r^2) \] \, , \label{eq:regcondN} \\
\Up & = & a \[ 1 + a r \vec{\al} \cdot \hat{X} + \Ord(r^2) \] \, , \label{eq:regcondUp} \\
U_a & = & \Ord(r^3)  \, , \label{eq:regcondUa} \\
\ga_{ab} & = & a^2 r^2 \[ 1 + 2 a r \vec{\al} \cdot \hat{X} \] {\rm diag}(1, \sin^2 \te)_{ab} + \Ord(r^4) \, , \label{eq:regcondgaab}
\eea
with the extra information that $a(t)$ is the scale factor in cosmological perturbation theory and \eqref{eq:rotX}. In particular, note that all the non-trivial information enters exclusively through the $\Up$ combination
\beq \label{eq:allonUp}
N = \Up + \Ord(r^2) \, , \hspace{1cm} \ga_{ab} = \Up^2 r^2 {\rm diag}(1, \sin^2 \te)_{ab} + \Ord(r^4) \, ,
\eeq
and that the observer dynamics $\{ \vec{\al}, \vec{\om} \}$ enter only through the combination $\vec{\al} \cdot \hat{X}[\vec{\om}]$. As for \eqref{eq:regcondUa}, it leads to the welcome feature of an observer 4-velocity \eqref{eq:nmu} with trivial angular part
\beq
n_o^{\mu} = \( a, a, 0, 0 \)^{\mu} \, ,
\eeq
which is also not the case in GLC in general \cite{Fanizza:2018tzp}. Observe next that, for a free-falling observer $\vec{\al} = 0$, the first two non-trivial orders of each metric component are the ones of the homogeneous and isotropic solution, which is clearly reminiscent of the Fermi normal coordinates, although in the present LC context. To make this relation precise, we trade the $w$ coordinate for $r := w - t$ and then the set $\{ r, \te^a \}$ for the corresponding Cartesian coordinates $\vec{x}$, to find that the line-element becomes
\beq
\ed s^2 = a^2(t) \[ - \ed t^2 + \ed \vec{x}^2 + \Ord(r^2) \] \, .
\eeq
These are the conformal Fermi normal coordinates \cite{Dai:2015rda}, which differ from the usual Fermi normal coordinates by the fact that $a(t)$ has been factored out. It therefore makes sense to refer to the LC coordinates supplemented by the extra conditions \eqref{eq:regcondN} to \eqref{eq:regcondgaab} as the (generalized) ``light-cone conformal Fermi'' coordinates, or ``LCCF'' for simplicity. The only leftover freedom is the choice of $N^{(n>2)}(t,\te)$ functions, which is still the bulk of the time-reparametrization freedom. In section \ref{sec:gaugeinv} we will see that these functions can be chosen such that the redshift, or some cosmological distance, is given by their background value to all orders in cosmological perturbation theory, or such that the equations of motion take a particularly natural form, analogous to the FLRW-longitudinal gauge.

\section{Geometrical preliminaries} \label{sec:prel}

\subsection{$d+1$ foliation} 

We consider a $D$-dimensional manifold $\cM$ with metric $g$ of arbitrary signature and let $d := D-1$. We then invoke a local coordinate system of the form $\{ y, x^i \}$, thus locally slicing (or ``foliating") $\cM$ in $y = {\rm const.}$ hypersurfaces which we denote by $\Si_y$. They all share the same topology $\Si_y \simeq \Si$, so we locally have $\cM \simeq \Rs \times \Si$. The line-element decomposes as follows
\beq \label{eq:ds2ADM}
\ed s^2 = s N^2 \ed y^2 + h_{ij} \( \ed x^i - s N^i \ed y \) \( \ed x^j - s N^j \ed y \) \, , \hspace{1cm} s = \pm 1 \, ,
\eeq
where $N$, $N^i$ and $h_{ij}$ are respectively known as the ``lapse function", ``shift vector" and ``$d$-metric", and $s$ allows us to consider both the time-like and space-like cases simultaneously.\footnote{The choice of putting an $s$ in front of $N^i$ is only conventional, as it can always be reabsorbed in $N^i$.} The lapse and shift form the unit normal vector to $\Si_y$
\beq \label{eq:nofalbe}
n := \frac{s g^{y\mu}}{\sqrt{s g^{yy}}}\, \pa_{\mu} \equiv N^{-1} \( \pa_y + s N^i \pa_i \) \, , \hspace{1cm} g (n,n) \equiv s \, .
\eeq
A useful expression for computations is
\beq \label{eq:nmudown}
n_{\mu} := g_{\mu\nu} n^{\nu} \equiv s N \de^y_{\mu} \, .
\eeq
To see that $n$ is indeed the normal vector to the $\Si_y$, note that the tangent space of the latter is generated by the $d$ vectors $\pa_i$ and
\beq
g (n, \pa_i) \sim \de_i^y \equiv 0 \, .
\eeq
We see that $N$ and $N^i$ capture the mismatch between the normal vector to $\Si_y$ and the vector $\pa_y$ which is parallel to motion along $y$, i.e. when the $x^i$ are held fixed (by definition of a partial derivative). Finally, to interpret $h_{ij}$, we first define 
\beq
h_{\mu}^{\nu} := \de_{\mu}^{\nu} - s n_{\mu} n^{\nu} \, ,
\eeq
which is the projector onto the tangent space of $\Si_y$
\beq
h^{\mu}_{\nu} n^{\nu} \equiv 0 \, , \hspace{1cm} h_{\mu}^{\ro} h_{\ro}^{\nu} \equiv h_{\mu}^{\nu} \, , \hspace{1cm} h_{\mu}^{\mu} \equiv d \, .
\eeq
The spatial components of $h_{\mu\nu}$ are the $d$-metric components $h_{ij}$, so the latter is the induced metric on $\Si_y$. Note also that the spatial components of $h^{\mu\nu}$, i.e. $h^{ij}$, coincide with the components of the inverse matrix of $h_{ij}$, so the notation is consistent. 

Now if we change the way we slice $\cM$, we change $n$. However, in this construction the slicing information is contained in the coordinate choice and, in particular, in the way the $y$ and $x^i$ coordinates are split. As a consequence, $n$ does {\it not} transform as a vector under all coordinate transformations. This is clear from the fact that its components are made of tensor components \eqref{eq:nofalbe}. We then see that it transforms as a vector 
\beq
n'^{\mu}(x') = \frac{\pa x'^{\mu}}{\pa x^{\nu}}\, n^{\nu}(x) \, ,
\eeq
only under the following subgroup of coordinate transformations
\beq \label{eq:SPCT}
y \to y'(y) \, , \hspace{1cm} x^i \to x'^i(y, x) \, ,
\eeq
in which case the metric components transform as follows
\bea 
N'(y',x') & = & \frac{\ed y}{\ed y'}\, N(y,x) \, , \nn \\
N'^i(y',x') & = & \frac{\ed y}{\ed y'} \[ \frac{\pa x'^i}{\pa x^j}\, N^j(y,x) - s \frac{\pa x'^i}{\pa y} \] \, , \label{eq:albegatrans} \\
h'^{ij}(y',x') & = & \frac{\pa x'^i}{\pa x^k} \frac{\pa x'^j}{\pa x^l}\, h^{kl}(y,x) \, . \nn
\eea
The second equation in \eqref{eq:SPCT} reparametrizes each slice $\Si_y$ independently, i.e. we perform a different $x^i$-coordinate transformation for each $y$ value. In contrast, the first transformation in \eqref{eq:SPCT} amounts to a reparametrization of $y$ that is the same for all $x^i$, meaning that we are just reparametrizing the slices
\beq
\Si_y \to \Si_{y'(y)} \, .
\eeq
Thus, \eqref{eq:SPCT} is the largest subgroup preserving the slicing of $\cM$, i.e. the $y = {\rm const.}$ hypersurfaces remain the same submanifolds of $\cM$, as one could have expected from the fact that $n$ is invariant. For this reason, we will refer to \eqref{eq:SPCT} as the ``slicing-preserving coordinate transformations" (SPCT). Consequently, any expression involving $n$ will only be invariant under this subgroup. Nevertheless, it is convenient to use $D$-dimensional notation anyway and keep referring to these objects as ``tensors". 

One can next note that $N$ and $N^i$ can be chosen arbitrarily, i.e. they are pure gauge variables, since any values can be obtained by performing a generic coordinate transformation on the simplest choice
\beq
N_* = 1 \, , \hspace{1cm} N_*^i = 0 \, ,
\eeq
which is known as the ``synchronous gauge" in the time-like case. Indeed, denoting the corresponding coordinates by $(y_*, x_*^i)$ and performing an arbitrary coordinate transformation to some $(y, x^i)$, we get that $N$ and $N^i$ are precisely the information of the Jacobian matrix between the two systems
\beq
\pa_y y_* = N \, , \hspace{1cm} \pa_y x^i_* = s N^j \pa_j x_*^i \, ,
\eeq
and these can be chosen arbitrarily indeed. On the other hand, if we start with an arbitrary coordinate system $(y,x^i)$ with lapse $N$ and $N^i$, we can perform the following SPCT
\beq
y \to y \, , \hspace{1cm} x^i \to x_*^i(y, x) \, ,
\eeq 
so that, using (\ref{eq:albegatrans}), the resulting shift is
\beq
N^i \to \frac{\pa x_*^i}{\pa x^j}\, N^j - s \, \frac{\pa x_*^i}{\pa y} = 0 \, .
\eeq
In contrast, one cannot obtain $N \to 1$ by using a SPCT in general, because the multiplicative factor $\ed y/ \ed y'$ in the transformation (\ref{eq:albegatrans}) does not depend on $x^i$. Thus, the information of the slicing lies exclusively in $N$, i.e. a choice of slicing amounts to a choice of $N$. On the other hand, $N^i$ controls how the points on a given slice $\Si_y$ are connected to the points of the next one $\Si_{y+\ed y}$, i.e. it controls the way the slices are ``glued" together.

\subsection{Gauss-Codazzi-Mainardi equations}

Having a $D$-dimensional tensor $h_{\mu\nu}$ representing the $d$-geometry of $\Si_y$ provides a straightforward way to express tensors on $\cM$ in terms of tensors on $\Si_y$. One first defines the projection operation onto $\Si_y$ 
\beq
(T^{\parallel})^{\mu_1 \dots \mu_m}_{\nu_1 \dots \nu_n} := h^{\mu_1}_{\ro_1} \dots h^{\mu_m}_{\ro_m} h^{\si_1}_{\nu_1} \dots h^{\si_n}_{\nu_n} T^{\ro_1 \dots \ro_m}_{\si_1 \dots \si_n} \, ,
\eeq
and refer to the tensors satisfying $T \equiv T^{\parallel}$, or equivalently $n \cdot T \equiv 0$, as ``tangent" (to $\Si_y$) tensors. Working with a tangent vector as an example $X^{\mu}$, we stress that one must be careful with the position of the indices, because $n_{\mu} X^{\mu} \equiv 0$ implies
\beq
X^y \equiv 0 \, , \hspace{1cm} {\rm but} \hspace{1cm} X_y \equiv - s N^i X_i \, .
\eeq
Nevertheless, the independent components lie in the spatial part and the two versions are consistently related by the $d$-metric
\beq
X_i \equiv g_{i\mu} X^{\mu} \equiv g_{iy} X^y + g_{ij} X^j \equiv h_{ij} X^j \, .
\eeq
Thus, as long as we focus on the purely spatial components of tangent $D$-tensors, their position is irrelevant and they transform covariantly under SPCTs
\beq
T'^{i_1 \dots i_m}_{j_1 \dots j_n}(y',x') = \frac{\pa x'^{i_1}}{\pa x^{k_1}} \dots \frac{\pa x'^{i_n}}{\pa x^{k_n}} \, \frac{\pa x^{l_1}}{\pa x'^{j_1}} \dots \frac{\pa x^{l_n}}{\pa x'^{j_n}}\, T^{k_1 \dots k_m}_{l_1 \dots l_n}(y,x) \, ,
\eeq
where $\pa x^j/ \pa x'^i$ is the inverse matrix of $\pa x'^i/\pa x^j$. It then turns out that the space of such tensors is naturally endowed with a unique ``tangent" covariant derivative $\na^{\parallel}_{\mu}$, defined by
\beq
\na^{\parallel} T := (\na T)^{\parallel} \, .
\eeq
Indeed, in full analogy with the $D$-dimensional case, this is the only derivation that is compatible with the induced metric
\beq
\na^{\parallel}_{\ro} h_{\mu\nu} \equiv 0 \, ,
\eeq
and has zero torsion
\beq
\[ \na^{\parallel}_{\mu}, \na^{\parallel}_{\nu} \] \ph \equiv 0 \, ,
\eeq
only now we also have the extra property
\beq
n^{\mu} \na^{\parallel}_{\mu} \equiv 0 \, .
\eeq
This allows one to implicitly define a Riemann tensor for $h_{\mu\nu}$ in the usual way
\beq \label{eq:Rperpdef}
\[ \na^{\parallel}_{\mu}, \na^{\parallel}_{\nu} \] X^{\ro} =: (\cR^{\parallel})^{\ro}_{\,\,\,\si\mu\nu}\, X^{\si} \, , \hspace{1cm} X \equiv X^{\parallel} \, .
\eeq
where $\cR^{\parallel}_{\mu\nu\ro\si}$ is explicitly tangent. One can then check that the spatial components of $\cR^{\parallel}_{\mu\nu\ro\si}$ are equal to the Riemann tensor built out of $h_{ij}$
\beq \label{eq:RperpRga}
\cR^{\parallel}_{ijkl} \equiv R_{ijkl}[h] \, .
\eeq
The $\cR^{\parallel}_{\mu\nu\ro\si}$ tensor is therefore the ``intrinsic'' curvature of $\Si_y$, as it knows nothing about how these hypersurfaces are curved in the $n$ direction, i.e. in the ambient $\cM$ space. This information is instead stored in the ``extrinsic'' curvature tensor\footnote{One also often finds the opposite sign convention for this definition.}
\beq \label{eq:Kdef}
K_{\mu\nu} := h_{\mu}^{\ro} \na_{\ro} n_{\nu} \equiv \frac{1}{2}\, \Lie_n h_{\mu\nu} \, ,
\eeq
where $\Lie_n$ is the Lie derivative in the $n$ direction 
\beq
\Lie_n h_{\mu\nu} := n^{\ro} \pa_{\ro} h_{\mu\nu} + h_{\ro\nu} \pa_{\mu} n^{\ro} + h_{\mu\ro} \pa_{\nu} n^{\ro} \, ,
\eeq
and the form $\sim \Lie_n h_{\mu\nu}$ is obtained using the specific expression (\ref{eq:nofalbe}). This tensor is symmetric and tangent
\beq
K_{\mu\nu} \equiv K_{\nu\mu} \, , \hspace{1cm} n^{\mu} K_{\mu\nu} \equiv 0 \, ,
\eeq
so all its independent information lies in $K_{ij}$. As in the case of the intrinsic curvature (\ref{eq:RperpRga}), here too the spatial components of (\ref{eq:Kdef}) provide the direct relation to $h_{ij}$
\beq \label{eq:Kofga}
K_{ij} \equiv \frac{1}{2N} \( \pa_y + s \Lie_N \) h_{ij} \, ,
\eeq
where now $\Lie_N$ is the Lie derivative in the $N^i$ direction on $\Si_y$
\beq
\Lie_N h_{ij} := N^k \pa_k h_{ij} + h_{kj} \pa_i N^k + h_{ik} \pa_j N^k \equiv D_i N_j + D_j N_i \, ,
\eeq
and $D_i$ is the covariant derivative made out of $h_{ij}$ on $\Si_y$. Note that the passage from \eqref{eq:Kdef} to \eqref{eq:Kofga} holds for any tangent $D$-tensor $T$
\beq
(\Lie_n T)^{i_1 \dots i_m}_{j_1 \dots j_n} \equiv N^{-1} \( \pa_y + s \Lie_N \) T^{i_1 \dots i_m}_{j_1 \dots j_n} \, ,
\eeq
because the terms containing derivatives of the lapse are proportional to a contraction of $T$ and $n$. As a result, when acting on $d$-tensors, the operator $N^{-1} \( \pa_y + s \Lie_N \)$ is a covariant derivation under SPCTs (\ref{eq:SPCT}). For instance, $K_{ij}$ transforms tensorially as $h_{ij}$
\beq
K'^{ij}(y',x') = \frac{\pa x'^i}{\pa x^k} \frac{\pa x'^j}{\pa x^l}\, K^{kl}(y,x) \, .
\eeq
One last identity that is needed is the derivative of $n$ along itself
\beq \label{eq:ngeo}
\na_n n_{\mu} \equiv -s\na_{\mu}^{\parallel} \log N \, ,
\eeq
found using the specific expression (\ref{eq:nofalbe}). This is consistent with the fact that all the geometric information of the slicing lies in $N$ alone. In particular, $N = 1$ is known as ``geodesic slicing". With \eqref{eq:ngeo} and \eqref{eq:Kdef} we can express the derivative of $n_{\mu}$ as
\beq
\na_{\mu} n_{\nu} \equiv \( h_{\mu}^{\ro} + s n_{\mu} n^{\nu} \) \na_{\ro} n_{\nu} \equiv K_{\mu\nu} - n_{\mu} \na_{\nu}^{\para} \log N \, . 
\eeq
We can now derive the Gauss-Codazzi-Mainardi equations. Expressing the right-hand side of (\ref{eq:Rperpdef}) in terms of $\na_{\mu}$ and using $n_{\nu} \na_{\mu} X^{\nu} \equiv -X^{\nu} \na_{\mu} n_{\nu}$, one obtains the Gauss-Codazzi equation
\beq \label{eq:GC}
h_{\mu}^{\al} h_{\nu}^{\be} h_{\ro}^{\ga} h_{\si}^{\de} R_{\al\be\ga\de}[g] \equiv \cR^{\parallel}_{\mu\nu\ro\si} - s \[ K_{\mu\ro} K_{\nu\si} -K_{\mu\si} K_{\nu\ro} \] \, ,
\eeq
which expresses the tangent part of the Riemann tensor of $\cM$ in terms of the intrinsic and extrinsic curvatures of $\Si_y$. The full Riemann tensor information is then found by also considering parallel components, i.e. the Codazzi-Mainardi equations
\bea
h_{\mu}^{\al} h_{\nu}^{\be} h_{\ro}^{\ga} n^{\de} R_{\al\be\ga\de}[g] & \equiv & \na^{\parallel}_{\mu} K_{\nu\ro} - \na^{\parallel}_{\nu} K_{\mu\ro} \, ,  \label{eq:CM1} \\
h_{\mu}^{\al} n^{\be} h_{\nu}^{\ga} n^{\de} R_{\al\be\ga\de}[g] & \equiv & - \Lie_n K_{\mu\nu} + K_{\mu\ro} K^{\ro}_{\nu} - s N^{-1} \na^{\parallel}_{\mu} \na^{\parallel}_{\nu} N   \, ,  \label{eq:CM2}
\eea
which are found using
\beq
R_{\al\be\ga\de} n^{\de} \equiv R_{\ga\de\al\be} n^{\de} \equiv \[ \na_{\al}, \na_{\be} \] n_{\ga} \, ,
\eeq
and expressing this in terms of $\na_n$ and $\na^{\parallel}_{\mu}$.

\section{General Relativity in light-cone coordinates} \label{sec:GRLC}

\subsection{$4 \to 3+1$} \label{sec:stdADM}

We now apply the foliation procedure described in the previous section to the case of interest, i.e. we pick $D = 4$ with a Lorentzian metric $g_{\mu\nu}$ and foliate the manifold with respect to the time-like direction $s = -1$. From now on we therefore denote the $y$ coordinate by ``$t$". The $n^{\mu}$ vector is thus time-like and can therefore be interpreted as the 4-velocity of a family of observers, as discussed in section \ref{sec:extra}. The energy, momentum and stress tensor measured by that family is then
\beq \label{eq:EPSdef}
E := n^{\mu} n^{\nu} T_{\mu\nu} \, , \hspace{1cm} P^{\mu} := - h^{\mu\nu} n^{\ro} T_{\nu\ro} \, , \hspace{1cm} S^{\mu\nu} := h^{\mu\ro} h^{\nu\si} T_{\ro\si} \, ,
\eeq
respectively, and the inverse decomposition can be compactly expressed as
\beq \label{eq:TmunuofEPS}
T_{\mu\nu} \ed x^{\mu} \ed x^{\nu} \equiv E \( N \ed t \)^2 - 2 P_i \( N \ed t \) \( \ed x^i + N^i \ed t \) + S_{ij} \( \ed x^i + N^i \ed t \) \( \ed x^j + N^j \ed t \) \, .
\eeq
The case where matter is a perfect fluid is treated in detail in the appendix \ref{app:PF}. Let us start by decomposing the Einstein equations. We first consider the non-trivial double trace of \eqref{eq:GC} to get the ``time-time" component of the Einstein tensor
\beq
n^{\mu} n^{\nu} G_{\mu\nu} \equiv \frac{1}{2} \[ \cR^{\parallel} + K^2 - K_{\mu\nu} K^{\mu\nu} \] \, .
\eeq
In terms of the fields on $\Si_t$, the corresponding component of the Einstein equation thus reads
\beq \label{eq:H}
E = \frac{1}{2} \[ R[h] + K^2 - K_{ij} K^{ij} \] \, ,
\eeq
where we use natural units $c = 8\pi G = 1$. Next, we consider the non-trivial trace of (\ref{eq:CM1}) to get the ``time-space" components
\beq
h^{\mu\nu} n^{\ro} G_{\nu\ro} = \na^{\parallel}_{\nu} K^{\nu\mu} - (\na^{\parallel})^{\mu} K \, ,
\eeq
so that the corresponding Einstein equation gives
\beq \label{eq:Hi}
P_i = D_i K - D_j K^j_i \, .
\eeq
Finally, the ``space-space" components of the Ricci tensor are found using (\ref{eq:GC}) and (\ref{eq:CM2})
\bea
h_{\mu}^{\ro} h_{\nu}^{\si} R_{\ro\si} & \equiv & g^{\al\be} h_{\mu}^{\ro} h_{\nu}^{\si} R_{\al\ro\be\si} \equiv h^{\al\be} h_{\mu}^{\ro} h_{\nu}^{\si} R_{\al\ro\be\si} - n^{\al} n^{\be} h_{\mu}^{\ro} h_{\nu}^{\si} R_{\al\ro\be\si}  \nn \\
 & \equiv & \Lie_n K_{\mu\nu}  - 2 K_{\mu\ro} K^{\ro}_{\nu}  + K_{\mu\nu} K + \cR^{\parallel}_{\mu\nu} - N^{-1} \na_{\mu}^{\parallel} \na_{\nu}^{\parallel} N \, ,  
\eea
so the corresponding components of the Einstein equation in the alternative form
\beq
R_{\mu\nu} = T_{\mu\nu} - \frac{1}{2}\, g_{\mu\nu} T \, ,
\eeq
read
\beq \label{eq:Kdot}
\( \pa_t - \Lie_N \) K_{ij} = N \[ 2 K_{ik} K_j^k - K_{ij} K - R_{ij}[h] + S_{ij} - \frac{1}{2}\, h_{ij} \( S - E \)  \] + D_i D_j N \, .
\eeq
Along with (\ref{eq:Kofga})
\beq \label{eq:qdot}
\( \pa_t - \Lie_N \) h_{ij} = 2 N K_{ij} \, ,
\eeq
these two equations form the dynamical part of the Einstein equations in first-order form if one considers $K_{ij}$ as independent. Equations \eqref{eq:H} and \eqref{eq:Hi} are then constraint equations that are consistently preserved through evolution if they hold on the initial conditions. This set of equations could have been equivalently obtained by considering the ADM equations \cite{Arnowitt:1962hi}, trading the conjugate momenta $\pi^{ij}$ for
\beq
K_{ij} \equiv \frac{1}{\sqrt{h}} \[ \pi_{ij} - \frac{1}{2}\, h_{ij} \pi \] \, ,
\eeq
and using the Hamiltonian constraint \eqref{eq:H} to modify the dynamical equation for $K_{ij}$. The form considered here was originally obtained by York \cite{York:1978gql} and is usually referred to as the ``standard ADM form" of the equations in the numerical relativity literature. As for the matter sector, we have the energy-momentum conservation equations $\na_{\mu} T^{\mu\nu} = 0$ that we can express in $3+1$ form, i.e. as evolution equations for $E$ and $P_i$
\bea
\( \pa_t - \Lie_N \) E & = & - D_i( N P^i ) - P^i D_i N - N \(  K E + K_{ij} S^{ij} \) \, ,  \label{eq:evolE} \\
\( \pa_t - \Lie_N \) P_i & = & - D_j ( N S^j_i ) - E D_i N - N K P_i   \, . \label{eq:evolPi}
\eea
Finally, note how the shift vector $N^i$ enters only through the combination $\pa_t - \Lie_N$ in all evolution equations \eqref{eq:Kdot}, \eqref{eq:qdot}, \eqref{eq:evolE} and \eqref{eq:evolPi}. The Lie derivative is the generator of diffeomorphisms, so an evolution equation of the form $\pa_t X = \Lie_N X + \dots$ implies that, at every infinitesimal time step, one can perform an arbitrary infinitesimal 3-diffeomorphism $x^i \to x^i - N^i(t,\vec{x})$. Therefore, this is how the freedom of performing spatial reparametrizations $x^i \to x'^i(t,\vec{x})$ manifests itself along time-evolution.

\subsection{$3+1 \to 2+1+1$}

Now that our equations are expressed on the $\Si_t$ manifolds, we further split $x^i \to \{ w, \te^a \}$, where $a \in \{1,2 \}$, thus foliating each $\Si_t$ into $w = {\rm const.}$ surfaces $S_{t,w}$. We will therefore now apply the Gauss-Codazzi-Mainardi formalism on $\Si_t$ with $s = 1$, thus obtaining a ``$2+1+1$" decomposition of our equations. Decomposing the line-element of $\Si_t$ in the ADM fashion \eqref{eq:ds2ADM}
\beq \label{eq:3metric}
\ed l^2 := h_{ij}\, \ed x^i \ed x^j = \Up^2 \ed w^2 + \ga_{ab} \( \ed \te^a - U^a \ed w \) \( \ed \te^b - U^b \ed w \) \, ,
\eeq 
the unit-normal vector to $S_{t,w}$ is
\beq
\nu := \frac{h^{wi}}{\sqrt{h^{ww}}}\, \pa_i \equiv \Up^{-1} \( \pa_w + U^a \pa_a \) \, , \hspace{1cm} h (\nu,\nu) \equiv 1 \, ,
\eeq
and the projector onto $S_{t,w}$ is
\beq
\ga_{ij} := h_{ij} - \nu_i \nu_j \, .
\eeq
Moreover, we choose the $S_{t,w}$ surfaces to have spherical topology, so that $w$ can be thought of as a radius on $\Si_t$, while the $\te^a$ are angles. We can next define the extrinsic curvature of $S_{t,w}$
\beq 
\cC_{ij} := \ga_i^k D_k \nu_j \equiv \frac{1}{2}\, \Lie_{\nu} \ga_{ij} \, ,
\eeq
so that, in full analogy with \eqref{eq:Kdef}, its angular components read
\beq \label{eq:Cabdef}
\cC_{ab} \equiv \frac{1}{2}\, \Up^{-1} \( \pa_w + \Lie_U \) \ga_{ab} \, ,
\eeq
where here $\Lie_U$ is the Lie derivative with respect to $U^a$ on $S_{t,w}$
\beq
\Lie_U \ga_{ab} \equiv U^c \pa_c \ga_{ab} + \ga_{cb} \pa_a U^c + \ga_{ac} \pa_b U^c \equiv \na_a U_b + \na_b U_a \, ,
\eeq
and $\na_a$ is the covariant derivative made out of $\ga_{ab}$ on $S_{t,w}$. The indices of tangent tensors to $S_{t,w}$ are displaced using $\ga_{ab}$. It is also convenient to extract the components of $K_{ij}$, $P_i$ and $S_{ij}$ under the $S_{t,w}$ foliation
\beq
\Te := \nu^i \nu^j K_{ij} \, , \hspace{1cm} A_i := - 2 \ga_i^j \nu^k K_{jk} \, , \hspace{1cm} \cK_{ij} := \ga_i^k \ga_j^l K_{kl} \, ,
\eeq
\beq \label{eq:Pidecomp}
\cP := \nu^i P_i \, , \hspace{1cm} \cP_i := \ga_i^j P_j \, , 
\eeq
and
\beq \label{eq:Sijdecomp}
\Si := \nu^i \nu^j S_{ij} \, , \hspace{1cm} \cS_i := \ga_i^j \nu^k S_{jk} \, , \hspace{1cm} \cS_{ij} := \ga_i^k \ga_j^l S_{kl} \, .
\eeq

\subsection{The light-cone shift vector} \label{sec:LCshift}

We are now in a position to describe the LC line-element \eqref{eq:GenLC} in this context. It simply amounts to the following choice of shift vector  
\beq \label{eq:LC}
N^i = - N \nu^i \, .
\eeq
With this, although $n$ is time-like, and thus $\Si_t$ is space-like, the vector $\pa_t$ is light-like
\beq
g(\pa_t , \pa_t) \equiv g_{tt} = 0 \, ,
\eeq
and radially oriented
\beq
\pa_t \equiv N n + N^i \pa_i = N \( n - \nu^i \pa_i \) \, .
\eeq
Thus, evolving along $t$, while keeping $w,\te^a$ constant, means moving along a light-like direction. Since the $S_{t,w}$ submanifolds have spherical topology, the subset
\beq
L_w := \bigcup_t S_{t,w} \, ,
\eeq
therefore forms a light-cone. Note also that the shift condition \eqref{eq:LC} is preserved only if both sides transform as 3-vectors, meaning that it breaks the reparametrization freedom $x^i \to \ti{x}^i(t,\vec{x})$ of the $\Si_t$ slices down to the SPCTs of the $S_{t,w}$ slices, which are nothing but the residual freedom \eqref{eq:RGT}. 

Finally, with the shift now being fixed to \eqref{eq:LC}, and equations \eqref{eq:Pidecomp} and \eqref{eq:Sijdecomp}, the energy-momentum tensor \eqref{eq:TmunuofEPS} reads
\bea 
T_{\mu\nu} \ed x^{\mu} \ed x^{\nu} & \equiv & E \( N \ed t \)^2 - 2 \( N \ed t \) \[ \cP \( \Up \ed w - N \ed t \) + \cP_a \( \ed \te^a - U^a \ed w \) \] + \Si \( \Up \ed w - N \ed t \)^2 \nn \\
 & & +\, 2 \cS_a \( \Up \ed w - N \ed t \) \( \ed \te^a - U^a \ed w \) + \cS_{ab} \( \ed \te^a - U^a \ed w \) \( \ed \te^b - U^b \ed w \)  \, . \label{eq:TmunuofEcPcS}
\eea

\subsection{The Einstein and energy-momentum conservation equations}   \label{sec:GReq}

We can now finally express the equations of GR in terms of the $2+1+1$ fields. We start by projecting \eqref{eq:qdot} to get
\bea
\dot{\Up} & = & \Up \[ N \Te - N' \] \, , \label{eq:dotUp} \\
\dot{U}^a & = & N \[ \Up A^a - \na^a \Up \] + \Up \na^a N \, , \label{eq:dotUa} \\
\dot{\ga}_{ab} & = & 2 N \( \cK_{ab} - \cC_{ab} \) \, , \label{eq:dotgab}
\eea
where the dot denotes $\pa_t$ and we have introduced the notation for tangent tensors to $S_{t,w}$
\beq
X' := \Lie_{\nu} X \equiv \Up^{-1} \( \pa_w + \Lie_U \) X \, ,
\eeq
e.g.
\beq
\cC_{ab} \equiv \frac{1}{2}\, \ga'_{ab} \, .
\eeq
From these equations we infer
\beq
\dot{\nu}^i = - \( N \Te - N' \) \nu^i + N \[ A^i - D^i_{\parallel} \log \Up \] + D^i_{\para} N \, , \hspace{1cm} \dot{\nu}_i = \( N \Te - N' \) \nu_i \, .
\eeq
We will also need the following relations
\bea
\nu^i \nu^j R_{ij}[h] & \equiv & - \, \cC' - \cC_{ab} \cC^{ab} - \Up^{-1} \na^2 \Up \, , \\
\ga_a^i \nu^j R_{ij}[h] & \equiv & \na_b \cC^b_a - \na_a \cC \, , \\
\ga_a^i \ga_b^j R_{ij}[h] & \equiv & - \cC'_{ab} + 2 \cC_{ac} \cC^c_b - \cC \cC_{ab} + \frac{1}{2}\,\ga_{ab} \cR - \Up^{-1} \na_a \na_b \Up \, , 
\eea
where $\cR \equiv R[\ga]$ is the Ricci scalar of $\ga_{ab}$ and we have used the fact that in two dimensions the Einstein tensor vanishes identically
\beq
\cR_{ab} \equiv \frac{1}{2}\, \ga_{ab} \cR \, . 
\eeq
These equations follow from the Gauss-Codazzi-Mainardi equations \eqref{eq:GC}, \eqref{eq:CM1} and \eqref{eq:CM2} applied to the 2+1 split $h_{ij} \to \ga_{ij} + \nu_i \nu_j$. Finally, we have the analogue of \eqref{eq:ngeo}
\beq 
D_{\nu} \nu_i \equiv - D_i^{\parallel} \log \Up \, .
\eeq
With \eqref{eq:Kdot} and the above equations we can now compute the following time-derivatives
\bea
\dot{\Te} & = & N \[ - \Te' + \cC' + \Up^{-1} \( \na^2 \Up + A^a \na_a \Up \) -\Te \[ \Te + \cK \]  - \frac{1}{2}\, A_a A^a + \cC_{ab} \cC^{ab} + \frac{1}{2} \( \Si - \cS + E \) \] \nn \\
 & & + \( \na_a \log \Up - A_a \) \na^a N + N'' \, , \label{eq:dotTe} \\
\nn \\
\dot{A}_a & = & N \[ - A'_a + 2 \( \cK_a^b - \Te \de_a^b \) \na_b \log \Up - \cK A_a  + 2 \( \na_b \cC_a^b - \na_a \cC - \cS_a \) \] \nn \\
 & & + \, 2 \[ \( \cC_a^b - \cK_a^b + \Te \de_a^b \) \na_b N - \na_a N' \]  \, ,  \label{eq:dotAa} \\
\nn \\
\dot{\cK}_{ab} & = & N \[ - \cK'_{ab} + \cC'_{ab} - \Te \cK_{ab} + 2 \cK_{ac} \cK^c_b - \cK_{ab} \cK - 2 \cC_{ac} \cC^c_b + \cC_{ab} \cC + \frac{1}{2}\, A_a A_b \right. \nn \\
 & & \left. \hspace{0.5cm} - A_{(a} \na_{b)} \log \Up + \Up^{-1} \na_a \na_b \Up + \cS_{ab} - \frac{1}{2}\, \ga_{ab} \( \cR + \Si + \cS - E \) \] \nn \\
 & & +\, \na_a \na_b N + A_{(a} \na_{b)} N + \cC_{ab} N' \, .  \label{eq:dotKab}
\eea
As for the constraint equations \eqref{eq:H} and \eqref{eq:Hi}
\bea
E & = & \frac{1}{2} \[ \cR - \cC^2 - \cC_{ab} \cC^{ab} + \( \cK + 2 \Te \) \cK - \cK_{ab} \cK^{ab} - \frac{1}{2}\, A_a A^a \] - \Up^{-1} \na^2 \Up - \cC' \, ,\label{eq:Econstr} \\
\nn \\
\cP & = & \cK^{ab} \cC_{ab} - \Te \cC  + \frac{1}{2}\, \na_a A^a + A^a \na_a \log \Up + \cK' \, , \label{eq:cPconstr} \\
\nn \\
\cP_a & = & \na_a \( \Te + \cK \) - \na_b \cK_a^b - \( \cK_a^b - \Te \de_a^b \) \na_b \log \Up + \frac{1}{2} \[ \, \cC A_a + A'_a \]  \, , \label{eq:cPaconstr}
\eea
and, finally, the energy-momentum conservation equations \eqref{eq:evolE} and \eqref{eq:evolPi}
\bea
\dot{E} & = & - N \[ E' + \cP' + \cC \cP + \na_a \cP^a + \cP^a \na_a \log \Up \right. \label{eq:dotE} \\
 & & \left. \hspace{0.7cm} + \( \Te + \cK \) E + \Te \Si - A^a \cS_a + \cK_{ab} \cS^{ab} \] - 2 \[ \cP N' + \cP^a \na_a N \] \, , \nn \\
\nn \\
\dot{\cP} & = & - N \[ \cP' + \Si' + \cC \Si + \na_a \cS^a + \( \cP^a + 2 \cS^a \) \na_a \log \Up \right.  \label{eq:dotcP} \\
 & & \left. \hspace{0.7cm} + \( 2 \Te + \cK \) \cP - A^a \cP_a - \cC_{ab} \cS^{ab} \] - \( E + \Si  \) N' + \( \cP^a - \cS^a \) \na_a N \, , \nn \\
\nn \\
\dot{\cP}_a & = & - N \[ \cP'_a + \cS'_a + \cC \cS_a + \na_b \cS^b_a + \[ \cS_a^b - \( \cP + \Si \) \de_a^b \] \na_b \log \Up + \( \Te + \cK \) \cP_a \] \label{eq:dotcPa} \\
 & & -\, \cS_a N' - \( E + \cP \) \na_a N - \cS_a^b \na_b N \, . \nn
\eea
Equations \eqref{eq:dotUp} to \eqref{eq:dotgab} and \eqref{eq:dotTe} to \eqref{eq:cPaconstr} form together the Einstein equations in LC form.

\section{Conformal fields}   \label{sec:confvar}

In this section we introduce a new set of fields, the ``conformal'' ones, in terms of which the regularity conditions and the derivation of perturbative equations become both simpler and more transparent. They are given by
\bea
\hat{N} & := & \Up^{-1} N \, , \\
\hat{\Up} & := & \log \Up \, , \label{eq:UphofUp} \\
\hat{U}_a & := & \Up^{-2} U_a \, , \label{eq:UhofU} \\ 
\hat{\ga}_{ab} & := & \Up^{-2} \ga_{ab} \, , \label{eq:gahofga} \\
\hat{\Te} & := & \Up \Te \, , \\
\hat{A}_a & := & A_a \, , \\
\hat{\cK}_{ab} & := & \Up^{-1} \( \cK_{ab} - \Te \ga_{ab} \) \, , \\
\hat{E} & := & \Up^2 E \, , \\
\hat{\cP} & := & \Up^2 \cP \, , \\
\hat{\cP}_a & := & \Up \cP_a \, , \\
\hat{\Si} & := & \Up^2 \Si \, , \\
\hat{\cS}_a & := & \Up \cS_a \, , \\
\hat{\cS}_{ab} & := & \cS_{ab} - \Si \ga_{ab} \, ,
\eea
and we use the convention of displacing the indices of hatted tensors with $\hat{\ga}_{ab}$. In terms of these fields the line-element \eqref{eq:GenLC} reads
\beq  \label{eq:LCds2conf}
\ed s^2 = e^{2\hat{\Up}} \[ - 2 \hat{N} \ed t \, \ed w + \ed w^2 + \hat{\ga}_{ab} \( \ed \te^a - \hat{U}^a \ed w \) \( \ed \te^b - \hat{U}^b \ed w \) \] \, ,
\eeq
so now $\hat{\Up}$ enters as an overall conformal factor, while it simply disappears from the energy-momentum tensor \eqref{eq:TmunuofEcPcS} 
\bea  
T_{\mu\nu} \ed x^{\mu} \ed x^{\nu} & \equiv & \hat{E} \( \hat{N} \ed t \)^2 - 2 \( \hat{N} \ed t \) \[ \hat{\cP} \( \ed w - \hat{N} \ed t \) + \hat{\cP}_a \( \ed \te^a - \hat{U}^a \ed w \) \] \nn \\
 & & +\, \hat{\Si} \[ \( \ed w - \hat{N} \ed t \)^2 + \hat{\ga}_{ab} \( \ed \te^a - \hat{U}^a \ed w \) \( \ed \te^b - \hat{U}^b \ed w \) \] \label{eq:TmunuofhEhcPhcS} \\
 & & +\, 2 \hat{\cS}_a \( \ed w - \hat{N} \ed t \) \( \ed \te^a - \hat{U}^a \ed w \) + \hat{\cS}_{ab} \( \ed \te^a - \hat{U}^a \ed w \) \( \ed \te^b - \hat{U}^b \ed w \)  \, . \nn
\eea
Finally, because of the relations \eqref{eq:allonUp}, the regularity conditions in the LCCF gauge \eqref{eq:regcondN} to \eqref{eq:regcondgaab} become extremely simple
\bea 
\hat{N} & = & 1 + \Ord(r^2) \, , \label{eq:regcondNc} \\
\hat{\Up} & = & \log a + a r \vec{\al} \cdot \hat{X} + \Ord(r^2)  \, , \label{eq:regcondUpc}  \\
\hat{U}_a & = & \Ord(r^3)  \, , \label{eq:regcondUac} \\
\hat{\ga}_{ab} & = & r^2 {\rm diag}(1, \sin^2 \te)_{ab} + \Ord(r^4) \, . \label{eq:regcondgaabc}
\eea
Let us now express the equations of the previous section in terms of the conformal fields. To that end, it is convenient to first write down some intermediary relations. We have
\beq
\cC_{ab} \equiv e^{\hat{\Up}} \[ \hat{\cC}_{ab} + \hat{\ga}_{ab} \hat{\Up}' \] \, , \hspace{1cm} \cC \equiv e^{-\hat{\Up}} \[ \hat{\cC} + 2 \hat{\Up}' \] \, , 
\eeq
where
\beq
\hat{\cC}_{ab} := \frac{1}{2}\, \hat{\ga}'_{ab} \, ,
\eeq
and from now on the prime is defined by
\beq
X' \to \( \pa_w + \Lie_U \) X \, ,
\eeq
i.e. without the $\Up^{-1}$ factor. Note also that 
\beq
\hat{U}^a := \hat{\ga}^{ab} \hat{U}_b \equiv U^a \, ,
\eeq
so that 
\beq
\Lie_U \equiv \Lie_{\hat{U}} \, .
\eeq
The Christoffels of $\ga_{ab}$ and $\hat{\ga}_{ab}$ are related by
\beq
\Ga^c_{ab} \equiv \hat{\Ga}^c_{ab} + \de^c_a \pa_b \hat{\Up} + \de^c_b \pa_a \hat{\Up} - \hat{\ga}_{ab} \hat{\ga}^{cd} \pa_d \hat{\Up} \, ,
\eeq
so the quantities of interest for us are
\bea
\cR & \equiv & e^{-2\hat{\Up}} \[ \hat{\cR} - 2\hat{\na}^2 \hat{\Up} \] \, , \\
\na_b \cC_a^b - \na_a \cC & \equiv & e^{-\hat{\Up}} \[ \hat{\na}_b \hat{\cC}_a^b - \hat{\na}_a \hat{\cC} + \hat{\cC}_a^b \hat{\na}_b \hat{\Up} - \hat{\na}_a \hat{\Up}' + \hat{\Up}' \hat{\na}_a \hat{\Up} \] \, , \\
\na_a \na_b f & \equiv & \hat{\na}_a \hat{\na}_b f - 2\hat{\na}_{(a} f \hat{\na}_{b)} \hat{\Up} + \hat{\ga}_{ab} \hat{\na}_c f \hat{\na}^c \hat{\Up}  \, , \\
\na^2 f & \equiv & e^{-2\hat{\Up}} \hat{\na}^2 f \, , 
\eea
where
$\hat{\cR}$ and $\hat{\na}$ are the Ricci tensor and covariant derivative of $\hat{\ga}_{ab}$ and $f$ is a scalar on $S_{t,w}$. The evolution equations become
\bea
\dot{\hat{\Up}} & = & \hat{N} \[ \hat{\Te} - \hat{\Up}' \] - \hat{N}' \, , \label{eq:dothUp} \\
\dot{\hat{U}}_a & = & \hat{N} \[ \hat{A}_a + 2 \( \hat{\cK}_a^b - \hat{\cC}_a^b \) \hat{U}_b \] + \hat{\na}_a \hat{N} + 2 \hat{U}_a \hat{N}' \, ,  \label{eq:dothUa} \\
\dot{\hat{\ga}}_{ab} & = & 2 \hat{N} \( \hat{\cK}_{ab} - \hat{\cC}_{ab} \) + 2 \hat{\ga}_{ab} \hat{N}'  \, , \label{eq:dothgab} \\
\dot{\hat{\Te}} & = & \hat{N} \[ - \hat{\Te}' + \hat{\cC}' + \hat{\Up}' \hat{\cC} - 2 \hat{\Te}^2 - \hat{\Te} \hat{\cK} + \hat{\na}^2 \hat{\Up} + 2 \hat{\na}_a \hat{\Up} \hat{\na}^a \hat{\Up} + 3 \hat{\Up}'' + \hat{\cC}_{ab} \hat{\cC}^{ab} - \frac{1}{2}\, \hat{A}_a \hat{A}^a \right. \nn \\
 & & \left. \hspace{0.6cm} +\, \frac{1}{2} \( \hat{E} - \hat{\Si} - \hat{\cS} \) \] + \( \hat{\na}_a \hat{\Up} - \hat{A}_a \) \hat{\na}^a \hat{N} - \( \hat{\Te} - \hat{\Up}' \) \hat{N}' + \hat{N}'' \, , \label{eq:dothTe}\\
\dot{\hat{A}}_a & = & \hat{N} \[ - \hat{A}'_a - \( \hat{\cK} + 2 \hat{\Te} \) \hat{A}_a + 2 \( \hat{\na}_b \hat{\cC}^b_a - \hat{\na}_a \hat{\cC} + 2 \hat{\cC}_a^b \hat{\na}_b \hat{\Up} - 2 \hat{\na}_a \hat{\Up}' + 2 \hat{\Up}' \hat{\na}_a \hat{\Up} - \hat{\cS}_a  \) \] \nn \\
 & & +\, 2 \( \hat{\cC}_a^b - \hat{\cK}_a^b \) \hat{\na}_b \hat{N} - 2 \hat{\na}_a \hat{N}' \, , \label{eq:dothAa} \\
\dot{\hat{\cK}}_{ab} & = & \hat{N} \[ - \hat{\cK}'_{ab} + \hat{\cC}'_{ab} - 2 \hat{\Te} \hat{\cK}_{ab} + 2 \hat{\Up}' \hat{\cC}_{ab} + 2 \hat{\cK}_{ac} \hat{\cK}^c_b - \hat{\cK}_{ab} \hat{\cK} - 2\hat{\cC}_{ac} \hat{\cC}^c_b + \hat{\cC} \hat{\cC}_{ab} + \frac{1}{2}\, \hat{A}_a \hat{A}_b \right. \nn \\
 & & \left. \hspace{0.6cm} +\, 2 \hat{\na}_a \hat{\na}_b \hat{\Up} - 2 \hat{\na}_a \hat{\Up} \hat{\na}_b \hat{\Up} + \hat{\ga}_{ab} \( -\, \hat{\cC}' - \hat{\cC}_{cd} \hat{\cC}^{cd} + \frac{1}{2}\, \hat{A}_c \hat{A}^c - \frac{1}{2}\, \hat{\cR} + 2 \hat{\Up}'^2 - 2 \hat{\Up}'' \) + \hat{\cS}_{ab}  \] \nn \\
 & & \hspace{0.6cm}  +\, \hat{N}' \[ \hat{\cK}_{ab} + \hat{\cC}_{ab} \] + \hat{A}_{(a} \hat{\na}_{b)} \hat{N} + \hat{\ga}_{ab} \hat{A}^c \hat{\na}_c \hat{N} + \hat{\na}_a \hat{\na}_b \hat{N} - \hat{\ga}_{ab} \hat{N}'' \, ,  \label{eq:dothKab}
\eea
the constraint equations become
\bea
\hat{E} & = & \frac{1}{2} \[ \hat{\cR} - \hat{\cK}_{ab} \hat{\cK}^{ab} + \hat{\cK}^2 - \hat{\cC}_{ab} \hat{\cC}^{ab} - \hat{\cC}^2 \] + 2 \hat{\cK} \hat{\Te} + 3 \hat{\Te}^2 - \hat{\cC}' - 2 \hat{\cC} \hat{\Up}'  \nn \\
 & & -\, 2 \hat{\na}^2 \hat{\Up} - \hat{\na}_a \hat{\Up} \hat{\na}^a \hat{\Up} - 2 \hat{\Up}'' - ( \hat{\Up}' )^2  - \frac{1}{4}\, \hat{A}_a \hat{A}^a \, , \label{eq:EconstrC} \\
\hat{\cP} & = & \hat{\cK}_{ab} \hat{\cC}^{ab} + \hat{\cK}' + 2 \hat{\Te}' + \frac{1}{2}\, \hat{\na}_a \hat{A}^a + \hat{A}^a \hat{\na}_a \hat{\Up} - 2 \hat{\Te} \hat{\Up}' \, , \label{eq:PconstrC} \\
\hat{\cP}_a & = & - \hat{\na}_b \hat{\cK}^b_a + \hat{\na}_a \( \hat{\cK} + 2 \hat{\Te} \) - 2 \( \hat{\cK}_a^b + \de_a^b \hat{\Te} \) \hat{\na}_b \hat{\Up} + \frac{1}{2}\, \hat{A}'_a + \frac{1}{2} \( \hat{\cC} + 2 \hat{\Up}' \) \hat{A}_a \, , \label{eq:PaconstrC}
\eea
and the energy-momentum conservation equations become
\bea
\dot{\hat{E}} & = & - \hat{N} \[ \hat{E}' + \hat{\cP}' + \( \hat{\cK} + \hat{\Te} \) \hat{E} + \( \hat{\cC} + 2 \hat{\Up}' \) \hat{\cP} + \hat{\na}_a \hat{\cP}^a + \( \hat{\cK} + 3 \hat{\Te} \) \hat{\Si} + \hat{\Te} \hat{\cS} \right. \nn \\
 & & \left. \hspace{1cm} +\, 2 \hat{\cP}^a \hat{\na}_a \hat{\Up} - \hat{A}^a \hat{\cS}^a + \hat{\cK}_{ab} \hat{\cS}^{ab} \] - 2 \( \hat{E} + \hat{\cP} \) \hat{N}' - 2 \hat{\cP}^a \hat{\na}_a \hat{N} \, , \label{eq:dotEC} \\
\dot{\hat{\cP}} & = & - \hat{N} \[ \hat{\cP}' + \( 2 \hat{\Te} + \hat{\cK} \) \hat{\cP} + \hat{\Si}'  + \hat{\Up}' \( \hat{E} - \hat{\Si} - \hat{S} \) + \hat{\na}_a \hat{\cS}^a - \hat{A}^a \hat{\cP}_a  + 2 \hat{\cS}^a \hat{\na}_a \hat{\Up} - \hat{\cC}_{ab} \hat{\cS}^{ab} \] \nn \\
 & & - \( \hat{E} + \hat{\Si} + 2 \hat{\cP} \) \hat{N}' + \( \hat{\cP}^a - \hat{\cS}^a \) \hat{\na}_a \hat{N} \, , \\
\dot{\hat{\cP}}_a & = & - \hat{N} \[ \hat{\cP}'_a + \hat{\cS}'_a + \( \hat{\cK} + 2 \hat{\Te} \) \hat{\cP}_a + \hat{\na}_b \hat{\cS}^b_a + \hat{\na}_a \Si + \( \hat{E} - \hat{\cS} - \hat{\Si} \) \hat{\na}_a \hat{\Up} + \( \hat{\cC} + 2 \hat{\Up}' \) \hat{\cS}_a + 2 \hat{\cS}_a^b \hat{\na}_b \hat{\Up} \] \nn \\
 & & - \( \hat{\cP}_a + \hat{\cS}_a \) \hat{N}' - \( \hat{E} + \hat{\Si} + \hat{\cP} \) \hat{\na}_a \hat{N} - \hat{\cS}_a^b \hat{\na}_b \hat{N} \, . \label{eq:dotcPa}
\eea

\section{Linear cosmological perturbation theory} \label{sec:linear}

\subsection{Background} 

In the LCCF coordinates the homogeneous and isotropic solution is given by  
\beq \label{eq:LCcompbg}
N = a(t) \, , \hspace{1cm} \Up = a(t) \, , \hspace{1cm} U_a = 0 \, , \hspace{1cm} \ga_{ab} = a^2(t)\, r_k^2(r) \, q_{ab}(\te)  \, ,
\eeq
where
\beq
r := w - t \, , \hspace{1cm} r_k(r) := \frac{\sin (\sqrt{k} \, r)}{\sqrt{k}} \, , \hspace{1cm} q_{ab} \, \ed \te^a \ed \te^b = \ed \te^2 + \sin^2\te \, \ed \vph^2 =: \ed \Om^2 \, ,
\eeq
with the latter indicating that we are also in the observational gauge \eqref{eq:standangparam}. The line-element \eqref{eq:GenLC} thus reads
\beq \label{eq:dsLCbg}
\ed s^2 = a^2 \[ - 2 \ed t \ed w + \ed w^2 + r_k^2(r) \, \ed \Om^2 \] \, ,
\eeq
so, in terms of $r$ we recover the FLRW form
\beq \label{eq:dsFLRWbg}
\ed s^2 = a^2 \[ - \ed t^2 + \ed r^2 + r_k^2(r) \, \ed \Om^2 \] \, , 
\eeq
meaning that $a$ is the scale factor, $t$ is conformal time, $r$ is the comoving distance to the observer and $k$ is the spatial curvature. Using \eqref{eq:dotUp}, \eqref{eq:dotUa}, \eqref{eq:dotgab} and \eqref{eq:Cabdef} we then find
\beq
\Te = a^{-1} \cH \, , \hspace{1cm} A_a = 0 \, , \hspace{1cm} \cK_{ab} = a^{-1} \cH \ga_{ab} \, , \hspace{1cm} \cC_{ab} = a^{-1} \cX \ga_{ab} \, , \hspace{1cm} \cR = \frac{2}{a^2 r_k^2} \, , 
\eeq
where we have defined the notation
\beq
\cH := \frac{\dot{a}}{a} \, , \hspace{1cm} \cX := \frac{r'_k}{r_k} \equiv \sqrt{k} \cot \( \sqrt{k}\, r \) \equiv \sqrt{\frac{1}{r_k^2} - k} \, ,  
\eeq
and the prime here reduces to differentiating with respect to $w$ only since $U_a$ vanishes. Note the useful identities
\beq
\dot{r}_k \equiv - \cX r_k \, , \hspace{1cm} r'_k \equiv \cX r_k \, , \hspace{1cm} \dot{\cX} \equiv \frac{1}{r_k^2} \, , \hspace{1cm} \cX' \equiv - \frac{1}{r_k^2}   \, .
\eeq
As for the matter sector
\beq
E = \ro(t) \, , \hspace{1cm}  \Si = p(t)  \, , \hspace{1cm} \cS_a = 0  \, , \hspace{1cm} \cS_{ab} = \ga_{ab} p(t) \, ,
\eeq
where here $\ro$ and $p$ are the background total energy density and pressure. Let us now consider the conformal fields defined in the previous section. Their background values are 
\beq
\hat{N} = 1 \, , \hspace{1cm} \hat{\Up} = \log a \, , \hspace{1cm} \hat{U}_a = 0  \, , \hspace{1cm} \hat{\ga}_{ab} = r_k^2 \, q_{ab} \, , 
\eeq
\beq
\hat{\Te} = \cH \, , \hspace{1cm} \hat{A}_a = 0 \, , \hspace{1cm} \hat{\cK}_{ab} = 0 \, , \hspace{1cm} \hat{\cC}_{ab} = \cX \hat{\ga}_{ab} \, , \hspace{1cm} \hat{\cR} = \frac{2}{r_k^2} \, ,
\eeq
and
\beq
\hat{E} = a^2 \ro =: \hat{\ro} \, , \hspace{1cm}  \hat{\Si} = a^2 p =: \hat{p}  \, , \hspace{1cm} \hat{\cS}_a = 0  \, , \hspace{1cm} \hat{\cS}_{ab} = 0 \, , 
\eeq
so their first advantage is the fact that half of the tensors are now zero, meaning a simpler derivation of perturbative equations. The second advantage will be the absence of explicit $a$ factors in the linearized equations, i.e. apart from the ones entering through $\cH$. Finally, the non-trivial equations at the background level are \eqref{eq:EconstrC}, \eqref{eq:dothTe} and \eqref{eq:dotEC}
\beq \label{eq:bgeq}
3 \( \cH^2 + k \) = \hat{\ro} \, , \hspace{1cm} 2\dot{\cH} + \cH^2 + k = - \hat{p} \, , \hspace{1cm} \dot{\hat{\ro}} = - \cH \( \hat{\ro} + 3 \hat{p} \) \, ,
\eeq
which are nothing but the Friedmann and energy conservation equations in terms of the conformal background variables, respectively.

\subsection{Perturbations} \label{sec:linpert}

Let us now consider fluctuations around the background solution. We will directly perform a 2d SVT decomposition with respect to the unit-sphere geometry $q_{ab}$. In particular, we will use ``$D_a$" to denote the corresponding covariant derivative and the indices of perturbative quantities will be displaced using $q_{ab}$. Since this is a two dimensional geometry, the pure-vector components can be expressed in terms of pseudo-scalars, i.e.
\beq
D_a h^a \equiv 0 \hspace{1cm} \Rightarrow \hspace{1cm} h_a \equiv \ti{D}_a \ti{h} \, ,
\eeq
where
\beq
\ti{D}_a := \ti{q}_a^{\,\,\,b} D_b \, , 
\eeq
and
\beq
\ti{q}_{ab} := \sqrt{q}\, \vep_{ab}  \, , 
\eeq
is the volume form on the unit-sphere. Therefore, a vector field decomposes as follows
\beq
V_a = D_a \bar{V} + \ti{D}_a \ti{V} \, ,
\eeq
and we will generically use a bar and a tilde to distinguish between the pure-scalar and pure-vector parts. For tensors of rank higher than one, we first note that $q_{ab}$ allows one to eliminate all pure-trace components, while $\ti{q}_{ab}$ allows one to eliminate any antisymmetric pair of indices through contraction, thus leaving only traceless and fully symmetric tensors to care about. One can then observe that pure-tensor components are zero because they obey as many constraints as their number of independent components. For instance, in the rank 2 tensor case
\beq
D_a h^{ab} \equiv 0 \, , \hspace{1cm} h_a^a \equiv 0 \hspace{1cm} \Rightarrow \hspace{1cm} h^{ab} \equiv 0 \, .
\eeq
Consequently, all non-trivial contributions come from the pure-scalar and pure-vector components. For instance, a traceless-symmetric rank 2 tensor field would first decompose as
\beq
T_{ab} = \bar{D}_{ab} \bar{T} + D_{(a} T_{b)} \, , \hspace{1cm} D_a T^a \equiv 0 \, ,  
\eeq
where
\beq
\bar{D}_{ab} := D_{(a} D_{b)} - \frac{1}{2}\, q_{ab} D^2 \, ,
\eeq
so that 
\beq
T_a = \ti{D}_a \ti{T} \, ,
\eeq
and thus
\beq
T_{ab} = D_{ab} \bar{T} + \ti{D}_{ab} \ti{T} \, ,
\eeq
where
\beq
\ti{D}_{ab} := \ti{D}_{(a} D_{b)} \equiv D_{(a} \ti{D}_{b)} \, .
\eeq
As a result, all perturbations are either scalars (barred) or pseudo-scalars (tilded) on the unit-sphere. Furthermore, if we decompose such a (pseudo-)scalar in the spherical harmonic basis on the sphere, then this amounts to decomposing the associated traceless-symmetric tensor of rank $s$ into spin-$s$ weighted spherical harmonics, since these are obtained by acting with $D_a$ and $\ti{D}_a$ on $Y_{lm}$.\footnote{More precisely, the spin weighted spherical harmonics arise if one works with a dyad $e_A^a$ on the sphere, i.e. $q_{ab} e_A^a e_B^b \equiv \de_{AB}$, and the corresponding derivatives in these directions $D_A := e_A^a D_a$. For more details see section 5.11 of \cite{Mitsou:2019nhj}.} The precise decomposition is 
\beq \label{eq:Ylmdecomp}
\ph(t,w,\te) = (-1)^s \sum_{l=s}^{\infty} \sqrt{\frac{\( l - s \)!}{\( l + s \)!}} \sum_{m=-l}^l \ph_{lm}(t,w)\, Y_{lm}(\te) \, ,
\eeq
where the $s$-dependent factor is inherited by the unit-normalization of spin-$s$ spherical harmonics and thus leads to a simple relation between the scalar product of the tensors and the ones of the $\ph_{lm}$ components, e.g. in the vector case we have
\bea
\int \ed \Om \, V_a V^a & = & \int \ed \Om \[ D_a \bar{V} D^a \bar{V} + D_a \ti{V} D^a \ti{V} \] \nn \\
 & = & -\sum_{l,l'=1}^{\infty} \frac{1}{\sqrt{l l' \( l+1 \) \( l' + 1 \)}} \sum_{m=-l}^l \sum_{m'=-l'}^{l'} \[ \bar{V}_{lm} \bar{V}_{l'm'} + \ti{V}_{lm} \ti{V}_{l'm'} \] \int \ed \Om \, Y_{lm} D^2 Y_{l'm'} \nn \\
 & = & \sum_{l,l'=1}^{\infty} \sqrt{\frac{l' \( l' + 1 \)}{l \( l+1 \)}} \sum_{m=-l}^l \sum_{m'=-l'}^{l'} \[ \bar{V}_{lm} \bar{V}_{l'm'} + \ti{V}_{lm} \ti{V}_{l'm'} \] \int \ed \Om \, Y_{lm} Y_{l'm'} \nn \\
 & = & \sum_{l=1}^{\infty} \sum_{m=-l}^l (-1)^m \[ \bar{V}_{lm} \bar{V}_{l,-m} + \ti{V}_{lm} \ti{V}_{l,-m} \]  \, .
\eea
From this it is also clear why the $l$ sum in \eqref{eq:Ylmdecomp} starts at $l = s$, because for $l < s$ the norm would be zero, meaning that $l$ is the total angular momentum. Thus, $\bar{V}_{lm}$ corresponds to an ``spin-1 $E$-mode", $\bar{T}_{lm}$ is a ``spin-2 $B$-mode", while their tilded counterparts correspond to the respective ``$B$-modes". The $E$/$B$-mode distinction reflects the even/odd behavior under parity, which is a symmetry of the background solution, so the two sectors will be decoupled at the linear level. Finally, since all fluctuations are (pseudo-)scalars on the sphere, the angular derivatives can only enter the linear equations through the Laplacian combination $D^2$. Thus, decomposing the fields as in \eqref{eq:Ylmdecomp} amounts to replacing
\beq \label{eq:toSphHar}
\ph(t,w,\te) \to (-1)^s \sqrt{\frac{\( l - s \)!}{\( l + s \)!}} \,\ph_{lm}(t,w) \, , \hspace{1cm} D^2 \to - l \( l + 1 \) \, , 
\eeq
in the equations. At the linear level all $lm$ modes are therefore decoupled, leaving us with a set of linear partial differential equations in the two variables $\{ t,w \}$. Including non-linear orders will bring couplings between different $lm$ values, but it will not change the fact that this is a system of 2-dimensional partial differential equations. 

We are now ready to express the conformal fields in terms of fluctuations around the background solution
\bea
\hat{N} & = & 1 + r_k^2 \psi \, , \label{eq:psidef} \\
\hat{\Up} & = & \log a + r_k^2 \ph \, , \label{eq:phperpdef} \\
\hat{U}_a & = & r_k^2 \( D_a \bar{u} + \ti{D}_a \ti{u} \)  \, ,  \label{eq:uhutdef} \\
\hat{\ga}_{ab} & = & r_k^2 \[ q_{ab} \( 1 + 2 r_k^2 \ch \) + 2 \( D_a D_b \bar{\chi} + \ti{D}_{ab} \ti{\chi} \) \] \, , \label{eq:phparachdef} \\ 
\hat{\Te} & = & \cH + r_k^2 \te \, , \\
\hat{A}_a & = & r_k^2 \[ D_a \bar{\al} + \ti{D}_a \ti{\al} \] \, , \\
\hat{\cK}_{ab} & = & r_k^2 \[ q_{ab} r_k^2 \ka + D_a D_b \bar{\ka} + \ti{D}_{ab} \ti{\ka} \] \, , \\
\hat{E} & = & \hat{\ro} \( 1 + r_k^2 \de \) \, , \label{eq:deltadef} \\
\hat{\cP} & = & \( \hat{\ro} + \hat{p} \) r_k^2 v \, , \\
\hat{\cP}_a & = & \( \hat{\ro} + \hat{p} \) r_k^2 \[ D_a \bar{v} + \ti{D}_a \ti{v} \] \, , \\
\hat{\Si} & = & \hat{p} + r_k^2 \si  \, , \\
\hat{\cS}_a & = & r_k^2 \[ D_a \bar{s} + \ti{D}_a \ti{s} \] \, , \\
\hat{\cS}_{ab} & = & r_k^2 \[ q_{ab} r_k^2 \pi + D_a D_b \bar{\pi} + \ti{D}_{ab} \ti{\pi} \] \, . \label{eq:Sabpertdef}
\eea
Several remarks are in order here. First, note that we chose to introduce the tensor $E$-modes through the operator $D_a D_b$, instead of $\bar{D}_{ab}$. The advantage is that the conformal 2-metric takes the form
\beq
\hat{\ga}_{ab} = r_k^2 \[ q_{ab} \( 1 + 2 r_k^2 \ch \) + D_a X_b + D_b X_a \] \, , \hspace{1cm} X_a := D_a \bar{\ch} + \ti{D}_a \ti{\ch} \, ,
\eeq
so that the spin-2 fields $\hat{\ch}$ and $\ti{\ch}$ enter as a $(t,w)$-dependent coordinate transformation, thus leaving $\ch$ to control the curvature $\cR$ of the $S_{t,w}$ surfaces. Second, note the non-trivial normalization of the fluctuations, i.e. the presence of the $r_k^2$ factors, even in front of scalars. This choice is ``natural'' in the sense that the corresponding linearized equations will have no explicit $r_k$ factors, just as they will have no explicit $a$ factors thanks to our use of the conformal variables. Instead, $a$ and $r_k$ will only enter through the combinations $\cH$, $\cX$ and the background Laplacian operator on the $S_{t,w}$ surfaces
\beq \label{eq:D2rk2}
\De^{(2)} := \frac{1}{r_k^2}\, D^2 \, .
\eeq
A priori, the only disadvantage is that the expansion in the distance to the observer starts at order $r^{-2}$, e.g.
\beq  \label{eq:phexp}
\ph(t,w,\te) = \sum_{n = -2}^{\infty} \ph^{(n)}(t,\te) \, r^n \, .
\eeq
Nevertheless, at least for the metric fluctuations, the first few orders are neutralized in a natural way by the regularity conditions in the LCCF gauge \eqref{eq:regcondNc} to \eqref{eq:regcondgaabc}. In the free-falling observer case $\vec{\al} = 0$ we have simply
\beq
\psi^{(n<0)}, \ph^{(n<0)}, \ch^{(n<0)} = 0 \, , 
\eeq
for the spin-0 fields,
\beq
\bar{u}^{(n<1)}, \ti{u}^{(n<1)} = 0 \, , 
\eeq
for the spin-1 fields, and
\beq
\bar{\ch}^{(n<2)}, \ti{\ch}^{(n<2)} = 0 \, , 
\eeq
for the spin-2 fields, i.e. the natural behavior that a spin-$s$ field starts at $\Ord(r^s)$. The more general case $\vec{\al} \neq 0$ can then be obtained by simply shifting $\ph$ 
\beq
\ph \to \ph + \frac{a}{r}\, \vec{\al} \cdot \hat{X}[\vec{\om}] \, .
\eeq
In fact, since $\vec{\al}$ and $\vec{\om}$ are not constrained by the equation of motion, this substitution can be performed at the end, i.e. at the level of the cosmological observable expressions. As for the matter sector, there are no regularity constraints other than the obvious ones
\beq
\hat{E}^{(0)} \equiv \hat{E}^{(0)}(t) \, , \hspace{0.5cm} \hat{\cP}^{(0)} \equiv \hat{\cP}^{(0)}(t) \, , \hspace{0.5cm} \hat{\Si}^{(0)} \equiv \hat{\Si}^{(0)}(t) \, , \hspace{0.5cm} \hat{P}_a^{(0)} \equiv 0 \, , \hspace{0.5cm} \hat{\cS}_a^{(0)} \equiv 0 \, , \hspace{0.5cm} \hat{\cS}_{ab}^{(n<1)} \equiv 0 \, , 
\eeq
meaning that the matter fluctuations start at $\Ord(r^{s-2})$. Moreover, note that the $\de$ field in \eqref{eq:deltadef} is the relative fluctuation of the conformal energy density $\hat{E}$, not of $E$ as is in the usual convention. On top of that, in the case where matter is a single perfect fluid (see appendix \ref{app:PF}) we have, at the linear level,
\beq \label{eq:PFpert}
\pi, \hat{\pi}, \ti{\pi}, \hat{s}, \ti{s} = 0 \, ,
\eeq
so these five fields control the linear anisotropic stress tensor. The speed of sound $c_s$ in terms of the canonical variables is implicitly defined by $\de \Si = c_s^2 \de E + \Ord(2)$, so in terms of the conformal fields $\hat{\Si}$, $\hat{E}$ we obtain, to linear order,
\beq \label{eq:PFcs}
\si = c_s^2 \hat{\ro} \( \de - 2 \ph \) + 2 \hat{p} \ph \, , 
\eeq   
where here $c_s(t)$ is the background speed of sound. The perfect fluid is then entirely determined by the data $\{ \de, v, \bar{v}, \ti{v} \}$ which can be evolved using the energy-momentum conservation equations. 

Before we close this subsection, let us perform an instructive exercise that reveals a subtlety in the difference between the LC and FLRW approaches. At the background level, the LC and FLRW metrics \eqref{eq:dsLCbg} and \eqref{eq:dsFLRWbg}, respectively, are related by the reparametrization $w \to r := w - t$. Since the latter is a finite transformation, it can only be implemented as a ``background'' coordinate transformation, under which the perturbation $\de g_{\mu\nu}$ transforms like a tensor, as opposed to the gauge transformations of cosmological perturbation theory. Taking the linearized LC line-element \eqref{eq:LCds2conf}
\bea
\frac{\ed s^2}{a^2} & = & - 2 \[ 1 + r_k^2 \( \psi + 2 \ph \) \] \ed t \ed w + \[ 1 + 2 r_k^2 \ph \] \ed w^2 - 2 r_k^2 \[ D_a \bar{u} + \ti{D}_a \ti{u} \] \ed w \ed \te^a  \nn \\ 
 & & +\, r_k^2 \[ q_{ab} \( 1 + 2 r_k^2\( \ch + \ph \) \) + 2 \( D_a D_b \bar{\ch} + \ti{D}_{ab} \ti{\ch} \) \] \ed \te^a \ed \te^b + \Ord(2) \, , \label{eq:LCfluct}
\eea
and replacing $w \to r + t$, we find
\bea
\frac{\ed s^2}{a^2} & = & - \[ 1 + 2 r_k^2 \( \psi + \ph \) \] \ed t^2 - 2 r_k^2 \psi \ed t \ed r - 2 r_k^2 \[ D_a \bar{u} + \ti{D}_a \ti{u} \] \ed t \ed \te^a + \[ 1 + 2 r_k^2 \ph \] \ed r^2 \label{eq:FLRWfluct} \\ 
 & & -\, 2 r_k^2 \[ D_a \bar{u} + \ti{D}_a \ti{u} \] \ed r \ed \te^a  + r_k^2 \[ q_{ab} \( 1 + 2 r_k^2 \( \ch + \ph \) \) + 2 \( D_a D_b \bar{\ch} + \ti{D}_{ab} \ti{\ch} \) \] \ed \te^a \ed \te^b + \Ord(2) \, . \nn
\eea
The former is indeed a perturbation of \eqref{eq:dsFLRWbg} and in the conformal Fermi normal coordinates \cite{Dai:2015rda} in particular. Its specificity is that the shift vector is fixed in terms of the rest of the fields. Focusing for simplicity on the flat case $k=0$, we can use Cartesian coordinates to find the relation 
\beq  \label{eq:g0iFLRW}
g_{0i} = - \frac{1}{2} \( g_{00} + \hat{x}^j \hat{x}^k g_{jk} \) \hat{x}^i + \( \de^{ij} - \hat{x}^i \hat{x}^j \) \hat{x}^k g_{jk} + \Ord(2) \, ,  
\eeq
where $r := |\vec{x}|$ and $\hat{x}^i := x^i/r$. Note that this line-element \eqref{eq:FLRWfluct} still possesses the nice property of canceling the line-of-sight integral terms in the expressions for observables, by construction. We thus see that one can implement that aspect of the LC idea through perturbation theory in FLRW coordinates. However, \eqref{eq:g0iFLRW} is clearly not part of the standard gauge choices in the literature. In particular, it explicitly privileges the radial direction through the $\hat{x}^i$ factors, which would then introduce derivatives in $k$-space and prevent a 3d SVT decomposition. Thus, the difference between the LC coordinates and the standard FLRW choices in the literature is not only the finite transformation $w \to r := w - t$, but also the hardwired dependence on the observer point on the LC side. Finally, in \cite{Fanizza:2020xtv} the authors showed how to obtain \eqref{eq:g0iFLRW} by considering general perturbation theory and imposing the GLC gauge at the perturbative level.

\subsection{Gauge-invariant variables} \label{sec:gaugeinv}

One clear lesson from cosmological perturbation theory in the FLRW coordinates is that the linearized equations become much simpler when expressed in terms of gauge-invariant field combinations. Moreover, this exercise provides a welcome consistency check of these equations by revealing their gauge invariance explicitly. Here the only freedom we have are the linearized time-reparametrizations that preserve the LCCF gauge
\beq \label{eq:lintrepa}
t \to t + r_k^2 T(t,w,\te) \, , \hspace{1cm} T^{(n<1)} = 0 \, .
\eeq
As for the other two residual symmetries \eqref{eq:RGT}, the light-cone reparametrizations have been fixed to the temporal gauge, i.e. the observer sits at $w = t$, while the $w$-dependent angular reparametrizations have been fixed to the non-rotating observational gauge (see section \ref{sec:extra}). In the case of the latter, however, this specification is not visible here, since we are using a covariant language to describe the $S_{t,w}$ submanifolds, i.e. we did not need to specify the functions $\hat{X}(\te)$, $q_{ab}(\te)$, or $D_a$ etc. It will therefore also be convenient to build invariant quantities under that symmetry as well, which we can express as follows
\beq \label{eq:linterepa}
\te^a \to \te^a + q^{ab} \[ D_b \bar{\Te}(w,\te) + \ti{D}_b \ti{\Te}(w,\te) \] \, 
\eeq
Implementing \eqref{eq:lintrepa} and \eqref{eq:linterepa} as a gauge transformation of the metric fluctuations, i.e. as a coordinate transformation in \eqref{eq:LCfluct} holding the background fixed, we obtain\footnote{Note that for the quantities of spin greater than zero, which therefore enter the original fields through angular derivatives, the transformations are defined only up to an arbitrary function of time only.} 
\bea
\psi & \to & \psi - \dot{T} - 2 \( T' + \cX T \) \, , \label{eq:psigt} \\
\ph & \to & \ph + T' + \( 2 \cX - \cH \) T  \, , \label{eq:phgt} \\
\bar{u} & \to & \bar{u} + \bar{\Te}' - T \, , \\
\ti{u} & \to & \ti{u} + \ti{\Te}' \, , \\
\ch & \to & \ch - T' - \cX T \, , \label{eq:chgt} \\
\bar{\ch} & \to & \bar{\ch} - \bar{\Te} \, , \label{eq:chbgt} \\
\ti{\ch} & \to & \ti{\ch} - \ti{\Te} \, . \label{eq:chtgt}
\eea
Note that, despite the fact that $T$ starts at $\Ord(r)$ (see \eqref{eq:lintrepa}), $T^{(1)}$ affects the zeroth order of the spin-0 fields $\{ \psi, \ph, \ch \}$ thanks to the derivatives and $\cX$ factors in \eqref{eq:psigt}, \eqref{eq:phgt} and \eqref{eq:chgt}, meaning that this symmetry allow us to impose one full condition on these three fields. Going back to the discussion at the end of section \ref{sec:LCCF}, we can choose this condition to simplify a given observable. For instance, for a source with 4-velocity $n^{\mu}$, so that the redshift is given by \eqref{eq:redshift}, setting $\ph = 0$ means that $1 + z = a_o/a$ to all orders in the perturbations. Another possibility is to trivialize the angular diameter distance, which is essentially controlled by the volume density $\ga^{1/4} \sim a r_k \[ 1 + r_k^2 \( \ph + \ch \) + \frac{1}{2}\, D^2 \bar{\ch} \]$ (see \cite{Fleury:2016htl}). 
  
As for the matter fluctuations, their transformation is obtained by proceeding similarly with the linearization of the energy-momentum tensor \eqref{eq:TmunuofhEhcPhcS} and we obtain
\bea
\de & \to & \de - \frac{\dot{\hat{\ro}}}{\hat{\ro}}\, T + 2 \( T' + 2 \cX T \) \, , \\
\si & \to & \si - \dot{\hat{p}} T + 2 \hat{p} \( T' + 2 \cX T \) \, , \\
v & \to & v + T' + 2 \cX T  \, , \\
\bar{v} & \to & \bar{v} + T \, , \\
\ti{v} & \to & \ti{v} \, , \\
\bar{s} & \to & \bar{s} \, , \\
\ti{s} & \to & \ti{s} \, , \\
\pi & \to & \pi \, , \\
\bar{\pi} & \to & \bar{\pi} \, , \\
\ti{\pi} & \to & \ti{\pi} \, , 
\eea
where $\hat{\ro}$ and $\hat{p}$ can be expressed in terms of $\cH$ and $k$ using the background equations \eqref{eq:bgeq}. To form gauge-invariant combinations, we can first form the invariant combinations under angular reparametrizations
\beq \label{eq:Udef}
\bar{U} := \bar{u} + \bar{\ch}' \, , \hspace{1cm} \ti{U} := \ti{u} + \ti{\ch}'  \, ,
\eeq
and with them the fully invariant combinations
\bea
\Psi & := & \psi - \dot{\bar{U}} - 2 \( \bar{U}' + \cX \bar{U} \)  \, , \\
\Phi & := & \ph + \bar{U}' + \( 2 \cX - \cH \) \bar{U}  \, , \\
\Om & := & \ch - \bar{U}' - \cX \bar{U} \, , \\
\de_U & := & \de - \frac{\dot{\hat{\ro}}}{\hat{\ro}}\, \bar{U} + 2 \( \bar{U}' + 2 \cX \bar{U} \)  \, , \\ 
\si_U & := & \si - \dot{\hat{p}} \bar{U} + 2 \hat{p} \( \bar{U}' + 2 \cX \bar{U} \) \, , \\
v_U & := & v + \bar{U}' + 2 \cX \bar{U}  \, , \\
\bar{v}_U & := & \bar{v} + \bar{U} \, .
\eea
From this we see that working with these variables is tantamount to working in the gauge 
\beq \label{eq:LClonggauge}
\bar{u} = - \bar{\ch}' \, ,
\eeq
since it means $\bar{U} = 0$, so that all of the above combinations reduce to the first term of each expression. This gauge is therefore reminiscent of the longitudinal gauge in the FLRW coordinates, so we will refer to it as the ``LCCF-longitudinal gauge''. As in the FLRW case, this gauge leads to the simplest form of the equations, i.e. the one with the minimal amount of time-derivatives and thus devoid of spurious degrees of freedom. Finally, the above transformations are consistent with the ones found in \cite{Fanizza:2020xtv}, where the authors work with general perturbations, if one further imposes the GLC conditions.

\subsection{Linearized equations of motion}

We can now linearize the equations of motion in conformal form, i.e. \eqref{eq:dothUp} to \eqref{eq:dotcPa}, around the cosmological background. The resulting expressions are given in appendix \ref{app:rawlin}. These are rather complicated, but become more transparent and amenable to resolution if we eliminate the ``momenta'' $ \{ \te,\bar{\al}, \ti{\al}, \ka, \bar{\ka}, \ti{\ka} \}$ through equations \eqref{eq:dotph} to \eqref{eq:dotcht} and express everything in terms of the gauge-invariant variables defined in the previous subsection. These are the equations we will display here. However, before we do so, two remarks are in order about the equations of appendix \ref{app:rawlin}. 

First, as mentioned previously, we observe that with the $a$ factors involved in the conformal variable definition, along with the $r_k$ factors in the normalization of the fluctuations, the linearized equations do not have any explicit dependence on $a$ and $r_k$. Rather, these quantities enter only through the combinations $\cH$ and $\cX$, respectively, and also the background angular Laplacian \eqref{eq:D2rk2} for $r_k$. Second, we see that the operator $\pa_t$ often enters in the combination 
\beq \label{eq:ringdef}
\ring{\ph} := \dot{\ph} + \ph' \, ,
\eeq
since this is the background form of the operator $\pa_t - \Lie_N$ in the $3+1$ equations. In our coordinates $\{ t, w, \te^a \}$ the operator $\pa_t$ denotes time derivation at constant light-cone $w$, so \eqref{eq:ringdef} is the time derivative at constant spatial radius $r \equiv w - t$. Indeed, trading the LC parametrization $\{ t, w \}$ for the FLRW one $\{ t, r \}$, the transformation of the partial derivatives
\beq
\ed t\, \pa_t + \ed w \, \pa_w \to \ed t\, \pa_t + \ed r \, \pa_r \, ,
\eeq
leads to
\beq \label{eq:wtor}
\dot{\ph} \to \ring{\ph} - \pa_r \ph \, , \hspace{1cm} \pa_w \ph \to \pa_r \ph  \, ,
\eeq
so the prime operator is the same, i.e. it is simply re-interpreted as the derivative with respect to $r$. Another convenient property of \eqref{eq:ringdef} is
\beq
\ring{r}_k \equiv \ring{\cX} \equiv 0 \, ,
\eeq
since these are functions of $r$ only. In what follows, we will use both $\dot{\ph}$ and $\ring{\ph}$, depending on what is the most convenient. We can now consider the linearized equations, starting with the gravitational evolution equations that are \eqref{eq:dotte} to \eqref{eq:dotka} 
\bea
\dot{\Psi}' + \( 2 \cX - \cH \) \( \dot{\Psi} - \Psi' \) - \( 2\dot{\cH} + \cH^2 - 4 \cH \cX + 2\cX^2 - 2 k \) \Psi  & & \label{eq:Phiperpddot} \\
+\, \rring{\Phi} + \cH \ring{\Phi} - 2 \Phi'' - 8 \cX \Phi' - 4 \( \cX^2 - k \) \Phi & & \nn \\
- \, \Om'' - 5 \cX \Om' - \( 3 \cX^2 - 3 k - \frac{1}{2}\, \De^{(2)} \) \Om & = &  - \frac{1}{2} \( \si_U + \De^{(2)} \bar{\pi} \) - \pi  \, , \nn \\
\nn\\
\dot{\Psi}' + \cX \dot{\Psi} - 2 \( \cX - \cH \) \Psi' + 2 \( \cH \cX + k \) \Psi - 2 \Phi'' - 6 \cX \Phi' + 4 k \Phi & & \label{eq:Phiparaddot0} \\
-\,\rring{\Om} - 2 \cH \ring{\Om} - \Om'' - 4 \cX \Om' + \( 4 k + \De^{(2)} \) \Om & = & -\pi \, , \nn \\
\nn\\
\dot{\Psi} - \Psi' + 2 \( \cH - \cX \) \Psi - 2 \( 2 \Phi + \Om \)' - 4 \cX \( \Phi + \Om \) + \bar{Q}' & = & 2 \bar{s}  \, , \label{eq:Psidot0} \\
\nn\\
\ddot{\ti{U}} + \dot{\ti{U}}' + 2 \cH \dot{\ti{U}} - \( 2 \cX^2 + 2 k + \De^{(2)} \) \ti{U} - \ti{Q}' & = & - 2 \ti{s}  \, , \label{eq:Utdot} \\
\nn\\
\bar{Q} - \Psi - 2\Phi & = & \bar{\pi} \, , \label{eq:chhdot} \\
\nn\\
\ti{Q} + \dot{\ti{U}} + 2 \( \cH - \cX \) \ti{U} & = & \ti{\pi} \, , \label{eq:chtdot}
\eea
where $\De^{(2)}$ is the background 2-dimensional Laplacian \eqref{eq:D2rk2} and 
\beq
\bar{Q} := \( \dot{\bar{\ch}} \)^{\circ} + 2 \cH \dot{\bar{\ch}} \, , \hspace{1cm} \ti{Q} := \( \dot{\ti{\ch}} \)^{\circ} + 2 \cH \dot{\ti{\ch}} \, .
\eeq
Next, we have the constraint equations \eqref{eq:constrde} to \eqref{eq:constrvt}
\bea
3 \( \cH^2 + k \) \de_U & = & 6 \cH \ring{\Phi} - 2 \[ \Phi'' + 6 \cX \Phi' + \( 6 \cX^2 - 2 k + \De^{(2)} \) \Phi \] \nn \\
 & & +\, 4 \cH \ring{\Om} - 2 \[ \Om'' + 7 \cX \Om' + \( 9 \cX^2 - k + \frac{1}{2}\, \De^{(2)} \) \Om \]  \nn \\
 & & +\, 2 \cH \[ \Psi' + \( 4 \cX - 3 \cH \) \Psi + \De^{(2)} \dot{\bar{\ch}} \] \, , \label{eq:constrdeU} \\
 \nn \\
2 \( - \dot{\cH} + \cH^2 + k \) v_U & = & 2 \[ \ring{\Phi}' + 2 \cX \ring{\Phi} - \cH \( \Phi' + 2 \cX \Phi \) \] + 2 \[ \ring{\Om}' + 3 \cX \ring{\Om} \] \nn \\
 & & -\, 2 \cH \Psi' - \( 4 \cH \cX + 2 k + \frac{1}{2} \, \De^{(2)} \) \Psi + \frac{1}{2}\, \De^{(2)} \[ \dot{\bar{\ch}}' + 2 \cX \dot{\bar{\ch}} \] \, , \label{eq:constrvU} \\
 \nn \\
2 \( - \dot{\cH} + \cH^2 + k \) \bar{v}_U & = & 2 \[ \ring{\Phi} - \cH \Phi \] + \ring{\Om} + \frac{1}{2}\, \Psi' - \( 2 \cH - \cX \) \Psi \nn \\
 & &  -\, \frac{1}{2}\, \dot{\bar{\ch}}'' - 2 \cX \dot{\bar{\ch}}' - \( \cX^2 + k \) \dot{\bar{\ch}}  \, , \label{eq:constrvbU} \\
\nn \\
2 \( - \dot{\cH} + \cH^2 + k \) \ti{v} & = & \frac{1}{2}\, \dot{\ti{U}}' + 2 \cX \dot{\ti{U}} - \( \cX^2 + k + \frac{1}{2}\, \De^{(2)} \) \ti{U} \nn \\
 & & -\, \frac{1}{2}\, \dot{\ti{\ch}}'' - 2 \cX \dot{\ti{\ch}}' - \( \cX^2 + k + \frac{1}{2}\, \De^{(2)} \) \dot{\ti{\ch}} \, . \label{eq:constrvtU}
\eea
and, finally, the energy-momentum conservation equations \eqref{eq:dotde} to \eqref{eq:dotvt}
\bea
3 \( \cH^2 + k \) \( \ring{\de}_U + \cH \de_U \) & = & 2 \( \dot{\cH} - \cH^2 - k \) \[ \( v_U + \Psi \)' + 4 \cX \( v_U + \Psi \) + \De^{(2)} \( \bar{v}_U + \dot{\bar{\ch}} \) + 2 \ring{\Om} \] \nn \\
 & & -\, 6 \dot{\cH} \( \cH \de_U - \ring{\Phi} \) - \cH \( 3 \si_U + 2 \pi + \De^{(2)} \bar{\pi}\)  \, ,  \label{eq:dotdeU} \\
 \nn \\
2 \( - \dot{\cH} + \cH^2 + k \) \( \ring{v}_U + 2 \cH v_U \) & = & 2 \( \ddot{\cH} - 2 \cH \dot{\cH} \) v_U - \si'_U - 2 \cX \si_U - \De^{(2)} \( \bar{s} - \cX \bar{\pi} \) + 2 \cX \pi  \\
 & & - \, 2 \( \dot{\cH} + 2 \cH^2 + 2 k \) \( \Phi' + 2 \cX \Phi \) + 2 \( \dot{\cH} - \cH^2 - k \) \( \Psi' + 2 \cX \Psi \)  \, ,  \nn \\
 \nn \\
2 \( - \dot{\cH} + \cH^2 + k \) \( \ring{\bar{v}}_U + 2 \cH \bar{v}_U \) & = & 2 \( \ddot{\cH} - 2 \cH \dot{\cH} \) \bar{v}_U - \bar{s}' - 4 \cX \bar{s} - \( \cX^2 + k + \De^{(2)} \) \bar{\pi} - \si_U - \pi  \nn \\
 & & - \, 2 \( \dot{\cH} + 2 \cH^2 + 2 k \) \Phi + 2 \( \dot{\cH} - \cH^2 - k \) \Psi  \, , \label{eq:dotvbU} \\
 \nn \\
2 \( - \dot{\cH} + \cH^2 + k \) \( \ring{\ti{v}} + 2 \cH \ti{v} \) & = & 2 \( \ddot{\cH} - 2 \cH \dot{\cH} \) \ti{v} - \ti{s}' - 4 \cX \ti{s} - \( \cX^2 + k + \frac{1}{2}\, \De^{(2)} \) \ti{\pi}  \, .
\eea
Although simpler than the ``raw'' equations in appendix \ref{app:rawlin}, the above equations still need some massaging before we can solve them, or at least interpret them physically. So let us start with the gravitational evolution equations and let us use \eqref{eq:chhdot} to eliminate $\bar{Q}$ from \eqref{eq:Psidot0}, obtaining
\beq \label{eq:Psidot}
\dot{\Psi} + 2 \( \cH - \cX \) \Psi - 2 \( \Phi + \Om \)' - 4 \cX \( \Phi + \Om \) = 2 \bar{s} - \bar{\pi}' \, .
\eeq
Using this equation to eliminate $\dot{\Psi}$ in \eqref{eq:Phiparaddot0} and then \eqref{eq:chtdot} to eliminate $\ti{Q}$ in \eqref{eq:Utdot}, we obtain the two completely decoupled wave equations
\bea  \label{eq:GW}
\bo \Om = - \pi - 2 \( \bar{s}' + \cX \bar{s} \) + \bar{\pi}'' + \cX \bar{\pi}'  \, , \hspace{1cm} \bo \ti{U} = 2 \ti{s} - \ti{\pi}' \, ,
\eea
where
\beq \label{eq:box}
\bo \ph := - \rring{\ph} - 2 \cH \ring{\ph} + \De^{(3)} \ph \, , \hspace{1cm} \De^{(3)} \ph := \ph'' + 2 \cX \ph' + \De^{(2)} \ph \, ,
\eeq
are the background scalar (conformal) d'Alembertian and Laplacian operators, respectively. Note that $\Om$ and $\ti{U}$ are the only fields on which angular derivatives act in the evolution equations, once $\bar{Q}$ and $\ti{Q}$ have been eliminated, and that they are both sourced by anisotropic stress components exclusively. These fields therefore describe the $E$ and $B$-mode gravitational wave excitations of GR, i.e. the degrees of freedom of the theory. Note that the first two non-trivial $r$-orders in each equation of \eqref{eq:GW} are constraints for $\Om^{(n<2)}$ and $\ti{U}^{(n<3)}$
\beq
D^2 \Om^{(0)} = - \pi^{(-2)}  \, , \hspace{1cm}  \( 2 + D^2 \) \Om^{(1)} = - \pi^{(-1)} - 2 \bar{s}^{(0)} \, , 
\eeq
and
\beq
\( 2 + D^2 \) \ti{U}^{(1)} = 2 \ti{s}^{(-1)} \, ,  \hspace{1cm} \( 6 + D^2 \) \ti{U}^{(2)} = 2 \ti{s}^{(0)} - \ti{\pi}^{(1)} \, ,
\eeq
so the degrees of freedom are in $\Om^{(n\geq 2)}$ and $\ti{U}^{(n\geq3)}$. Assuming $\Om$ and $\ti{U}$ solved, we are now left with the four fields $\{ \Psi, \Phi, \bar{\ch}, \ti{\ch} \}$, which do not carry any degrees of freedom. From the gravitational evolution equations we see that, a priori, we need to provide the initial conditions of seven fields $\{ \Psi, \Phi, \dot{\Phi}, \bar{\ch}, \ti{\ch}, \dot{\bar{\ch}}, \dot{\ti{\ch}} \}$. The initial conditions of $\{ \bar{\ch}, \ti{\ch} \}$ are pure-gauge, since they can be chosen arbitrarily through the light-cone-dependent angular reparametrizations \eqref{eq:chbgt} and \eqref{eq:chtgt}. Here this freedom is fixed to the non-rotating observational gauge, which implies a definite initial value for $\{ \bar{\ch}, \ti{\ch} \}$. However, this information is not required in order to evolve the equations, because these fields appear only through their time-derivatives $\{ \dot{\bar{\ch}}, \dot{\ti{\ch}} \}$, precisely because these are invariant under \eqref{eq:chbgt} and \eqref{eq:chtgt}. Let us now see how the other five initial conditions $\{ \Psi, \Phi, \dot{\Phi}, \dot{\bar{\ch}}, \dot{\ti{\ch}} \}$ are actually determined. We start with \eqref{eq:constrvtU}, which can be written as a Laplacian equation for $r_k \dot{\ti{\ch}}$
\beq
\frac{1}{r_k} \( \De^{(3)} + 3 k \) \[ r_k \( \dot{\ti{\ch}} + \ti{U} \) \] = \ring{\ti{U}}' + 4 \cX \ring{\ti{U}} - 4 \( - \dot{\cH} + \cH^2 + k \) \ti{v} \, ,
\eeq
thus determining that field in terms of the already solved $\ti{U}$ and $\ti{v}$. Next using \eqref{eq:constrvbU} to eliminate $\ring{\Phi}$ in \eqref{eq:constrdeU}, we obtain a Laplacian equation for $r_k^2 \Phi$
\bea
\frac{1}{r_k^2} \( \De^{(3)} - 3 \cH^2 \) \( r_k^2 \Phi \) & = & \frac{1}{4}\, \cH \( \Psi' + 10 \cX \Psi \) + \cH \[ \frac{3}{4}\, \dot{\bar{\ch}}'' + 3 \cX \dot{\bar{\ch}}' + \( \frac{3}{2} \, \cX^2 + \frac{3}{2} \,k + \De^{(2)} \) \dot{\bar{\ch}} \] \nn \\
 & & +\,  \frac{1}{2}\, \cH \ring{\Om} - \Om'' - 7 \cX \Om' - \( 9 \cX^2 - k + \frac{1}{2}\, \De^{(2)} \) \Om  \\
 & &  -\, \frac{3}{2} \( \cH^2 + k \) \de_U + 3 \cH \( - \dot{\cH} + \cH^2 + k \) \bar{v}_U  \, , \nn
\eea
which is the analogue of the usual Poisson equation in this context and therefore allows us to express $\Phi$ in terms of $\{ \Psi, \dot{\bar{\ch}}, \Om \}$ and matter fields. Taking the ring derivative of this expression and using \eqref{eq:chhdot}, \eqref{eq:Psidot}, \eqref{eq:GW}, \eqref{eq:dotdeU} and \eqref{eq:dotvbU} to eliminate $\{ \dot{\bar{\ch}}^{\circ}, \ring{\Psi}, \rring{\Om}, \ring{\de}_U, \ring{\bar{v}}_U \}$, we obtain an equation that allows to similarly eliminate $\dot{\Phi}$ in terms of $\{ \Psi, \dot{\bar{\ch}}, \Om \}$ and matter. We are finally left with $\{ \Psi, \dot{\bar{\ch}} \}$, which are determined through \eqref{eq:constrvU} and \eqref{eq:constrvbU}. We have therefore used all of the constraint equations and there are no degrees of freedom left indeed.

\subsection{Road-map to analytical solutions} \label{sec:roadmap}

The analytical solutions for specific types of matter will be given in future work, but let us conclude this section by providing some general remarks regarding that procedure. In the previous subsection we saw that we can rearrange the equations such that the operators acting on the fields are the familiar d'Alembertian $\bo$ and Laplacian $\De^{(3)}$ in \eqref{eq:box}, respectively. The central ingredient for solving such equations are the eigenfunctions $f_{\om,l}(r)$ of the Laplacian in spherical harmonic space
\beq \label{eq:deffnul}
f''_{\om,l}(r) + 2 \cX(r)\, f'_{\om,l}(r) - \frac{l \( l + 1 \)}{r_k^2(r)}\, f_{\om,l}(r) = - \om^2 f_{\om,l}(r) \, .
\eeq
Since this set of functions forms a basis, one can then use them to decompose the spatial dependence of every field
\beq \label{eq:gensphFourier}
\ph(t,w,\te) \sim \sum_{l=s}^{\infty} \sum_{m=-l}^l Y_{lm}(\te) \int \om^2 \ed \om \[ \ph^+_{\om,lm}(t)\, f^+_{\om,l}\( w-t \) + \ph^-_{\om,lm}(t)\, f^-_{\om,l}\( w-t \) \] \, ,
\eeq
where we distinguish between the eigenfunctions that are regular ($+$) and diverging ($-$) at $r = 0$. The precise behavior is obtained by expanding around $r = 0$ and using \eqref{eq:deffnul}
\beq
f^+_{\om,l}(r) \sim r^l + \Ord(r^{l+1}) \, ,  \hspace{1cm} f^-_{\om,l}(r) \sim r^{-l-1} + \Ord(r^{-l}) \, .
\eeq 
For the gravitational fields $\ph^-_{\om,lm} = 0$, thanks to the regularity conditions in the LCCF gauge, but since the matter fluctuations start at $\Ord(r^{-2})$, we have to consider $f^-_{\om,0}$ and $f^-_{\om,1}$ in that sector. As an example, decomposing $\Om$ as in \eqref{eq:gensphFourier} turns \eqref{eq:GW} into the ordinary differential equation
\beq
\ddot{\Om}^+_{\om,lm} + 2 \cH \dot{\Om}^+_{\om,lm} + \om^2 \Om^+_{\om,lm} = {\rm matter} \, .
\eeq 
The domain of the eigenvalues $\om$, i.e. the spectrum, and the eigenfunctions $f_{\om,l}(r)$ depend on the value of the spatial curvature $k$. In the flat case we have $\om \in \Rs^*$ and the spherical Bessel functions
\beq
f^+_{\om,l}(r) = j_l \( \om r \) \, ,  \hspace{1cm} f^-_{\om,l}(r) = y_l \( \om r \) \, ,  
\eeq
so that \eqref{eq:gensphFourier} becomes the spherical Fourier decomposition. In the open case $k < 0$ we also have $\om \in \Rs^+$, but it is convenient to use the parametrization $\om^2 = - k \nu \( 2 - \nu \)$, so that the IR region $\om^2 < -k$ is probed by $\nu \in [0,1]$, while the UV region $\om^2 > -k$ is probed by complex numbers of the form $\nu \in 1 + i \be$, where $\be \in \Rs^+$. The advantage is a simpler description of the eigenfunctions
\bea
f^+_{\nu,l}(r) & = & \frac{{}_2 F_1 \[ \frac{\nu + l}{2}, \frac{\nu + l + 1}{2}; \frac{1}{2} + l + 1; \tanh^2 \( \sqrt{-k}\, r \) \] }{\cosh^{\nu} \( \sqrt{-k}\,r \)} \, \tanh^l \( \sqrt{-k}\, r \) \, , \\
f^-_{\nu,l}(r) & = & \frac{{}_2 F_1 \[ \frac{\nu - l - 1}{2}, \frac{\nu - l}{2}; \frac{1}{2} - l; \tanh^2 \( \sqrt{-k}\, r \) \] }{\cosh^{\nu} \( \sqrt{-k}\,r \)} \, \tanh^{-l-1} \( \sqrt{-k}\, r \) \, , 
\eea
where ${}_2 F_1$ is the hypergeometric function. Finally, in the closed case $k > 0$ we have a discrete spectrum $\om^2 = k n \( 2 + n \)$, with $n \in \mathbb{N}$, and
\beq
f^+_{n,l}(r) = \frac{Q_{n+1/2}^{l+1/2}\( \cos(\sqrt{k}\,r) \) }{\sqrt{\sin(\sqrt{k}r)}}   \, ,  \hspace{1cm} f^-_{n,l}(r) = \frac{P_{n+1/2}^{l+1/2}\( \cos(\sqrt{k}\,r) \) }{\sqrt{\sin(\sqrt{k}r)}}   \, ,
\eeq
where $P$ and $Q$ are the Legendre functions of the first and second kind, respectively. One must then also replace the $\om$ integral with a discrete sum over $n \in [l, \infty[$ in \eqref{eq:gensphFourier}, which is then the hyperspherical harmonics decomposition.

\section{Statistics and the cosmological principle} \label{sec:statistics}

We now discuss the issue of imposing statistical isotropy {\it and} homogeneity with perturbation theory. Let $\ph(t,w,\te)$ be some field fluctuation and let us promote it to a stochastic field, working directly with the spherical harmonic components $\ph_{lm}(t,w)$. To determine the statistics of that field, it suffices to compute its $N$-point correlation functions at a given reference time $t_0$, say at the initial condition surface, since the time-evolution equations then allow one to compute all other unequal-time correlations. We are thus led to consider functions of the form $\bra \ph_{l_1 m_1}(t_0,w_1) \dots \ph_{l_N m_N}(t_0,w_N) \ket$. As shown in \cite{Mitsou:2019ocs}, and in particular equation (50) therein, statistical isotropy implies that the $l_k, m_k$ dependence can be reduced from $2N$ to $2N-3$ indices as follows
\beq
\bra \ph_{l_1 m_1}(t_0,w_1) \dots \ph_{l_N m_N}(t_0,w_N) \ket = \sum_{L_k} W_{m_1 \dots m_N}^{l_1 \dots l_N|L_1 \dots L_{N-3}}\, C_{l_1 \dots l_N|L_1 \dots L_{N-3}}(t_0; w_1, \dots, w_N) \, ,
\eeq    
where the $W^{\dots}_{\dots}$ coefficients are a generalization of the $3-j$ Wigner symbols given in equations (42) and (45) of \cite{Mitsou:2019ocs}. In particular, in the case of primary interest that is $N = 2$, we have $W_{m_1m_2}^{l_1l_2} \propto \de^{l_1 l_2} \de_{m_1,-m_2}$, so the information reduces to a spherical power spectrum of the form $C_l(t_0;w,w')$. We next note that the variables $w_k$ are the radial distances to the observer world-line on the $t = t_0$ hypersurface, up to an overall constant shift since $r_k := w_k - t_0$. One can therefore use the basis functions provided in section \ref{sec:roadmap} to obtain a full spectrum (in 3d). For instance, following \eqref{eq:gensphFourier} in the flat case $k=0$ we would have
\beq \label{eq:phlmbessel}
\ph_{lm}(t_0,w) = \int \om^2 \ed \om \[ \ph^+_{\om,lm}(t_0)\, f^+_{\om,l}\( w - t_0 \) + \ph^-_{\om,lm}(t_0)\, f^-_{\om,l}\( w-t_0 \) \] \, ,
\eeq 
which therefore leads to spectra of the form $C_{l_1 \dots l_N|L_1 \dots L_{N-3}}(t_0; \om_1, \dots, \om_N)$ and their analogues for the $k \neq 0$ cases. Since these quantities are defined on a constant time hypersurface, although in spherical coordinates, we can now impose statistical homogeneity in the same way we do in the FLRW approach with spherical Fourier decomposition (see for instance \cite{Yoo:2019qsl}). Coming back to the most relevant case $N = 2$, statistical homogeneity is the statement \beq
\bra \ph^{\pm}_{\om,lm}(t_0)\, \ph^{\pm}_{\om',lm}(t_0) \ket \propto \de \( \om - \om' \) \, , 
\eeq
and leads to the form
\beq
C^{\pm}_l(t_0;\om,\om') = \de \( \om - \om' \) P(t_0;\om) \, ,
\eeq
i.e. the usual single-entry power spectrum function. Thus, despite the fact that the LC coordinates privilege the observer world-line, one can still implement the cosmological principle in the statistics. 

The quantities $C_{l_1 \dots l_N|L_1 \dots L_{N-3}}(t_0; \om_1, \dots, \om_N)$ form the initial conditions when $\ph$ is one of the degrees of freedom of the theory. The other case where one would need to consider $N$-point spectra is when $\ph$ is a field combination that is an observable. Fixing the gauge freedom within LCCF to the trivial redshift $1 + z = a_o/a(t)$ to all orders in perturbation theory (see section \ref{sec:gaugeinv}), the quantity of interest would then be 
\beq \label{eq:Cobs}
C^{\rm obs}_{l_1 \dots l_N|L_1 \dots L_{N-3}}(z_1, \dots z_N) := \bra \ph_{l_1 m_1}(t(z_1),w_o) \dots \ph_{l_N m_N}(t(z_N),w_o) \ket \, .
\eeq
Indeed, this is a function relating observables and can therefore be compared to observations through appropriate estimators \cite{Mitsou:2019ocs}, i.e. this is the output information. The right-hand side of \eqref{eq:Cobs} can be obtained by expressing $\ph_{lm}(t,w)$ in terms of the initial conditions $\ph_{lm}(t_0,w)$ through the LC evolution equations and thus the input data $C_{l_1 \dots l_N|L_1 \dots L_{N-3}}(t_0; w_1, \dots, w_N)$. The whole point of the LC coordinates is to make this relation as straightforward as possible.

\section{Conclusion}  \label{sec:conclusion}

In this paper we have laid the foundations for a novel approach to analytical cosmological computations. The basic idea is simple: to relate in the most direct way the ``input data'' that are field spectra $C_{l_1 \dots l_N|L_1 \dots L_{N-3}}(t_0; \om_1, \dots, \om_N)$ on some initial hypersurface $t = t_0$, to the ``output data'' that are the angular spectra of observables in the sky $C^{\rm obs}_{l_1 \dots l_N|L_1 \dots L_{N-3}}(z_1, \dots, z_N)$. Our formalism is built upon the existing concept of light-cone adapted coordinates \cite{ObsCoord, Gasperini:2011us} and in particular the case where one of the foliations is made of space-like hypersurfaces \cite{Gasperini:2011us}. 

We have first extended the known example of such coordinates (GLC) to the largest possible family that exhibits the same advantageous properties: the fact that the input and output data lie on constant coordinate hypersurfaces and the fact that light-propagation towards the observer is trivial. In particular, the second property implies that expressions for observables contain no line-of-sight integrals, thus simplifying the aforementioned input-output map. Moreover, the generalization of GLC allowed us to identify convenient alternatives, such as the ``light-cone conformal Fermi coordinates'', which trivialize the gravitational field in the vicinity of the observer, and a further gauge fixing which trivializes the redshift observable. 

We have next considered the minimal set of equations in the context of relativistic cosmology, i.e. the Einstein and energy-momentum conservation equations, and expressed them in the LC coordinates in a geometrically controlled way. We have then provided all the required structure for setting up cosmological perturbation theory and we have worked out all relevant equations at the linear level. In particular, we identified the field combinations that are invariant under the remaining gauge symmetries in the LC family and verified that all our linearized equations are consistently gauge-invariant. Finally, we have also clarified how one can do statistics with the corresponding stochastic fluctuations without sacrificing the usual assumption of statistical homogeneity.

\acknowledgments

We are grateful to Giovanni Marozzi, Gabriele Veneziano and the participants of the second practitioner's workshop on relativistic effects in cosmology (Zurich, 2019) for useful discussions, to the referees for helpful suggestions, and especially to Fulvio Scaccabarozzi for collaboration at an early stage of this project. EM and JY are supported by a Consolidator Grant of the European Research Council (ERC-2015-CoG grant 680886), GF is supported by Fundaç\~ao para a Ci\^encia e a Tecnologia under the program {\it Stimulus} with the grant no. CEECIND/04399/2017/CP1387/CT0026 and NG and JY are supported by the Swiss National Science Foundation.

\appendix

\section{Regularity conditions} \label{app:obscoord}

Here we derive the non-trivial regularity conditions on the LC metric, i.e. equations \eqref{eq:conscond1nt} to \eqref{eq:conscondgaap3}. We employ the same procedure as in \cite{Fanizza:2018tzp}, but with an expansion at constant $t$ instead of constant $w$, and also generalize to arbitrary observer dynamics. We start by simply assuming that space-time is regular in the vicinity of the observer, so that there exists a coordinate system $(T,\vec{X})$ such that \cite{Misner:1974qy}
\beq \label{eq:gFC}
\ed s^2(T,\vec{X}) = - \[ 1 + 2 \vec{A}(T) \cdot \vec{X} \] \ed T^2 + 2 \[ \vec{\Om}(T) \times \vec{X} \] \cdot \ed T\, \ed \vec{X} + \ed \vec{X}^2 + \Ord(X^2) \, ,
\eeq
so that $T$ is the proper time of the observer, situated at $\vec{X} = 0$, $\vec{A}$ is their acceleration 3-vector and $\vec{\Om}$ is their spin 3-vector. The higher-order terms $\Ord(X^2)$ depend on the curvature tensor at the observer \cite{Ni:1978zz} and therefore on the specific space-time under consideration. Setting $\vec{A} = \vec{\Om} = 0$, one recovers the Fermi normal coordinates \cite{Manasse:1963zz}. 

We will now relate the coordinate system \eqref{eq:gFC} to the LC system and, in doing so, obtain the regularity conditions on the LC metric components. We first expand the coordinate transformation functions around the observer as in \eqref{eq:gexp}
\bea 
T(t,w,\te) & = & T^{(0)}(t) + T^{(1)} \( t, \te \) r + T^{(2)} \( t, \te \) r^2 + \dots \, , \label{eq:Texp} \\
\vec{X}(t,w,\te) & = & \vec{X}^{(1)} \( t, \te \) r + \vec{X}^{(2)} \( t, \te \) r^2 + \dots \, , \label{eq:TXexp}
\eea
where we remind that $r := w - t$. Inserting these two equations in \eqref{eq:gFC}, we obtain 
\bea
g_{tt} & = & - \[ 1 + 2 \vec{A}(T) \cdot \vec{X} \] ( \pa_t T )^2 + 2 \[ \vec{\Om}(T) \times \vec{X} \] \cdot \pa_t T\, \pa_t \vec{X} + ( \pa_t \vec{X} )^2 + \Ord[\( w - t \)^2] \, ,  \\
g_{tw} & = & - \[ 1 + 2 \vec{A}(T) \cdot \vec{X} \] \pa_t T \pa_w T + \[ \vec{\Om}(T) \times \vec{X} \] \cdot \( \pa_t T\, \pa_w \vec{X} + \pa_w T\, \pa_t \vec{X} \) + \pa_t \vec{X} \cdot \pa_w \vec{X} + \Ord[\( w - t \)^2] \, , \nn \\
g_{ta} & = & - \[ 1 + 2 \vec{A}(T) \cdot \vec{X} \] \pa_t T \pa_a T + \[ \vec{\Om}(T) \times \vec{X} \] \cdot \( \pa_t T\, \pa_a \vec{X} + \pa_a T\, \pa_t \vec{X} \) + \pa_t \vec{X} \cdot \pa_a \vec{X} + \Ord[\( w - t \)^3] \, , \nn \\
g_{ww} & = & - \[ 1 + 2 \vec{A}(T) \cdot \vec{X} \] ( \pa_w T )^2 + 2 \[ \vec{\Om}(T) \times \vec{X} \] \cdot \pa_w T\, \pa_w \vec{X} + ( \pa_w \vec{X} )^2 + \Ord[\( w - t \)^2] \, , \nn \\
g_{wa} & = & - \[ 1 + 2 \vec{A}(T) \cdot \vec{X} \] \pa_w T \pa_a T + \[ \vec{\Om}(T) \times \vec{X} \] \cdot \( \pa_w T\, \pa_a \vec{X} + \pa_a T\, \pa_w \vec{X} \) + \pa_w \vec{X} \cdot \pa_a \vec{X} + \Ord[\( w - t \)^3] \, , \nn \\
g_{ab} & = & - \[ 1 + 2 \vec{A}(T) \cdot \vec{X} \] \pa_a T \pa_b T + 2 \[ \vec{\Om}(T) \times \vec{X} \] \cdot \pa_{(a} T\, \pa_{b)} \vec{X} + \pa_a \vec{X} \cdot \pa_b \vec{X} + \Ord[\( w - t \)^4] \, . \nn
\eea
Next, we insert the expansions \eqref{eq:Texp} and \eqref{eq:TXexp}, using \eqref{eq:gexp} and only the obvious conditions \eqref{eq:U0ga0ga10} and \eqref{eq:consconds0}. We will consider only the first two non-trivial orders in each series, so that the resulting equations depend exclusively on the terms appearing in \eqref{eq:gFC} and are therefore independent of the specific space-time under consideration. The leading order equations are
\bea
0 & = & \( \dot{T}^{(0)} - T^{(1)} \)^2 - R^2   \, , \label{eq:tt0} \\
0 & = & \( \dot{T}^{(0)} - T^{(1)} \) \pa_a T^{(1)} + R \pa_a R  \, , \label{eq:ta1} \\
a^2 & = & \( \dot{T}^{(0)} - T^{(1)} \) T^{(1)} + R^2 \, , \label{eq:tw0} \\
a^2 + \ga_{(2)}^{ab} U^{(1)}_a U^{(1)}_b & = & - T_{(1)}^2 + R^2 \, , \label{eq:ww0} \\
U_a^{(1)} & = & \frac{1}{2} \, \pa_a \[ T_{(1)}^2 - R^2 \] \, , \label{eq:wa1} \\
\ga_{ab}^{(2)} & = & - \pa_a T^{(1)} \pa_b T^{(1)} + \pa_a \vec{X}^{(1)} \cdot \pa_b \vec{X}^{(1)} \, , \label{eq:ab2}
\eea
where 
\beq
R^2 := \vec{X}^{(1)} \cdot \vec{X}^{(1)} \, .
\eeq
Combining \eqref{eq:tt0} and \eqref{eq:tw0} we find
\beq \label{eq:N0Y0}
a^2 = \dot{T}^{(0)} \( \dot{T}^{(0)} - T^{(1)} \) \, ,
\eeq 
which therefore implies that 
\beq
T^{(1)}(t,\te) = T^{(1)}(t) \, .
\eeq
Inserting this in \eqref{eq:tt0}, we then find
\beq
R = \dot{T}^{(0)} - T^{(1)} \, , \hspace{1cm} \Rightarrow \hspace{1cm} R(t,\te) = R(t) \, ,
\eeq
meaning that the vectors $\vec{X}^{(1)}(t,\te)$ generate a sphere with radius $R(t)$. We can therefore decompose them as follows
\beq
\vec{X}^{(1)}(t,\te) \equiv R(t) \, \hat{X}(t,\te) \, , \hspace{1cm} \hat{X} \cdot \hat{X} \equiv 1 \, ,
\eeq
where the $\hat{X}$ allow us to construct the unit-sphere 2-metric $q_{ab}$ and volume 2-form $\ti{q}_{ab}$, defined in \eqref{eq:qtiqdef}. With this \eqref{eq:wa1} becomes $U_a^{(1)} = 0$, so that \eqref{eq:tt0}, \eqref{eq:tw0} and \eqref{eq:ww0} become 
\beq \label{eq:T10}
T^{(1)} = 0 \, ,  \hspace{1cm} a = R = \dot{T}^{(0)} \, ,
\eeq
and, finally, \eqref{eq:ab2} is $\ga_{ab}^{(2)} = a^2 q_{ab}$, i.e. we have obtained \eqref{eq:conscond1nt}. 

Now that all of the leading order equations are solved, we can consider the next-to-leading order. The corresponding expressions will contain $\dot{\hat{X}}$, so we must first express this quantity in a convenient way. We decompose $\dot{\hat{X}}$ in the basis $\{ \hat{X}, \pa_a \hat{X} \}$, but since $\hat{X} \cdot \hat{X} \equiv 1$, we have $\hat{X} \cdot \dot{\hat{X}} \equiv 0$ and thus
\beq
\dot{\hat{X}} \equiv a W^a \pa_a \hat{X} \, ,
\eeq
where $W^a(t,\te)$ is an angular velocity vector field on the unit-sphere. It can be recovered through
\beq \label{eq:Vadef}
W_a \equiv a^{-1} \dot{\hat{X}} \cdot \pa_a \hat{X} \, ,
\eeq
where we use $q_{ab}$ to displace angular indices. Defining the corresponding covariant derivative $D_a$, satisfying in particular
\beq \label{eq:DDXiden}
D_a D_b \hat{X} \equiv - q_{ab} \hat{X} \, ,
\eeq
and taking the divergence of \eqref{eq:Vadef}, we find
\beq
D_a W^a \sim D_a \hat{X} \cdot D^a \dot{\hat{X}} - 2 \hat{X} \cdot \dot{\hat{X}} \equiv \pa_t \[ \frac{1}{2}\, D_a \hat{X} \cdot D^a \hat{X} - \hat{X} \cdot \hat{X} \] \equiv 0 \, , 
\eeq
meaning that $W^a$ is a pure curl
\beq
W^a \equiv \ti{q}^{ab} \pa_b W \, .
\eeq
We can now consider the next-to-leading order equations. By decomposing $\vec{X}^{(2)}$ in the basis $\{ \hat{X}, \pa_a \hat{X} \}$
\beq
\vec{X}_{(2)} \equiv R_{(2)} \hat{X} + X_{(2)}^a \pa_a \hat{X} \, ,
\eeq
using the identities
\beq \label{eq:Xiden1}
\hat{X} \times \pa_a \hat{X} \equiv \ti{q}_a^{\,\,\,b} \pa_b \hat{X}   \, , \hspace{1cm} \pa_a \hat{X} \times \pa_b \hat{X} \equiv \ti{q}_{ab} \hat{X} \, ,
\eeq
and taking into account the leading-order equations, we obtain
\bea
0 & = & \dot{a} + a^2 \al - 2 \( R^{(2)} + T^{(2)} \) \, , \label{eq:tt1} \\
0 & = & - a^2 \ti{q}_a^{\,\,\,b} \pa_b \( W + \om \) + \pa_a \( R^{(2)} + T^{(2)} \) + X^{(2)}_a \, , \label{eq:ta2} \\
N^{(1)} + \Up^{(1)} & = & - \dot{a} + 2 T^{(2)} + 4 R^{(2)} \, , \label{eq:tw1} \\
\Up^{(1)} & = & 2 R^{(2)} \, , \label{eq:ww1} \\
U_a^{(2)} & = & - a \( \pa_a R^{(2)} + X^{(2)}_a \)  \, , \label{eq:wa2} \\
\ga_{ab}^{(3)} & = & 2 a \[ q_{ab} R^{(2)} + D_{(a} X_{b)}^{(2)} \] \, , \label{eq:ab3}
\eea
where
\beq
\al := \vec{\al} \cdot \hat{X}  \, , \hspace{1cm} \om = \vec{\om} \cdot \hat{X}   \, ,
\eeq
and
\beq
\vec{\al}(t) := \vec{A}(T^{(0)}(t)) \, , \hspace{1cm} \vec{\om}(t) := \vec{\Om}(T^{(0)}(t)) \, .
\eeq
We start by using \eqref{eq:tt1} to eliminate $R^{(2)}$ and decompose $V_a$ and $X^{(2)}_a$ harmonically on the 2-sphere (see section \ref{sec:linpert} for details)
\beq
X_a^{(2)} \equiv \pa_a X^{(2)} + \ti{q}_a^{\,\,\,b} \pa_b \ti{X}^{(2)} \, ,  \hspace{1cm}  X_a^{(2)} \equiv \pa_a X^{(2)} + \ti{q}_a^{\,\,\,b} \pa_b \ti{X}^{(2)} \, ,  
\eeq
so that equations \eqref{eq:ta2} to \eqref{eq:ab3} lead to
\bea
X^{(2)} & = & - \frac{1}{2}\, a^2 \al \, , \\
\ti{X}^{(2)} & = & a^2 \( W + \om \) \, , \\
N^{(1)} & = & a^2 \al \, , \\
\Up^{(1)} & = & \dot{a} + a^2 \al - 2 T^{(2)} \, , \\
U_a^{(2)} & = & a \pa_a T^{(2)} - a^3 \ti{q}_a^{\,\,\,b} \pa_b \( W + \om \) \, , \\
\ga_{ab}^{(3)} & = & a \[ \dot{a} + 2 a^2 \al - 2 T^{(2)} \] q_{ab} + 2 a^3 \ti{q}_{(a}^{\,\,\,\,c} D_{b)} \pa_c W \, .   
\eea
This leads to equations \eqref{eq:conscondN1} to \eqref{eq:conscondgaap3} if we trade $T^{(2)}$ for
\beq \label{eq:ZofT2}
Z := a^{-2} \[ \frac{1}{2}\, \dot{a} - T^{(2)} \] \, .
\eeq
To conclude this appendix, let us compare our derivation with the procedure employed in \cite{Fanizza:2018tzp}, where the authors focus on the GLC case $N = 1$ and also use the temporal gauge \eqref{eq:tempgauge}. There one considers an expansion of the form \eqref{eq:gexp}, but with coefficients that are functions of $(w,\te)$ instead of $(t,\te)$, i.e. it is an expansion in $w - t$ at constant $w$, instead of constant $t$. This is fine, since $w = t$ at the observer, but leads to a different expression of the regularity conditions than the ones obtained here. The reasons for choosing our approach instead are the following. First, being an expansion at constant space-like distance to the observer, instead of light-like as in \cite{Fanizza:2018tzp}, it is the same kind of expansion as in the generalized Fermi coordinates \eqref{eq:gFC}, thus leading to a more intuitive understanding of the mapping between the two systems. Second, and most importantly, our approach is more suited to cosmology, where $\Up$ is the scale factor $a(t)$ up to fluctuations. Given the temporal gauge \eqref{eq:tempgauge}, or \eqref{eq:consconds0}, in our approach we have the simple result $\Up_o(t) = N_o(t) \sim a(t)$, while in the approach of \cite{Fanizza:2018tzp} the condition $N = 1$ implies $\Up_o(w) = 1$, meaning that the $a(t)$ information is disseminated in the expansion in $t$.

\section{Perfect fluid matter $3+1$ decomposition} \label{app:PF}

Here we consider the case where the matter sector is made of a single perfect fluid. We can therefore express the energy-momentum tensor as 
\beq \label{eq:Trestdecomp}
T_{\mu\nu} = \ro \, V_{\mu} V_{\nu} + p \( g_{\mu\nu} + V_{\mu} V_{\nu} \)  \, , \hspace{1cm} V_{\mu} V^{\mu} \equiv -1 \, ,
\eeq
where $V^{\mu}$ is the 4-velocity, while $\ro$ and $p$ are the rest-frame energy density and pressure, i.e. as measured by an observer family with 4-velocity $V^{\mu}$, as opposed to $E$ and $S$ which are measured by the canonical observers with 4-velocity $n^{\mu}$. In order to relate the rest-frame and canonical quantities, it is useful to define 
\beq \label{eq:vWdef}
v^i := N^{-1} \( \frac{V^i}{V^0} - \frac{n^i}{n^0} \) \equiv N^{-1} \( \frac{V^i}{V^0} + N^i \) \, , \hspace{1cm} W := -n_{\mu} V^{\mu} \equiv \[ 1 - h_{ij} v^i v^j \]^{-1/2} \, ,
\eeq
that are the $3$-velocity of matter measured by the canonical observer $n^{\mu}$ and the generalization of the Lorentz factor, respectively, 
\beq
h^{i\mu} V_{\mu} \equiv W v^i \, . 
\eeq
Plugging \eqref{eq:Trestdecomp} inside \eqref{eq:EPSdef} we then find  
\bea
E & = & W^2 \( \ro + p \) - p \, , \\
P^i & = & W^2 \( \ro + p \) v^i  \, , \\
S^{ij} & = & W^2 \( \ro + p \) v^i v^j + p h^{ij} \, ,
\eea
so that, in particular, 
\beq \label{eq:SijPF}
S^{ij} = p h^{ij} + \frac{P^i P^j}{E + p} \, .
\eeq
We can also invert the energy and velocity relations to get  
\beq \label{eq:roE}
\ro = E - \frac{P_i P^i}{E + p} \, , \hspace{1cm} v^i = \frac{P^i}{E + p} \, .
\eeq
Eqs. \eqref{eq:evolE} and \eqref{eq:evolPi} are four equations for five independent variables $E$, $P^i$ and $p$, thanks to \eqref{eq:SijPF}, so the equation of state that relates the rest-frame scalars $p \equiv p(\ro(E,P))$ closes the system.

As explained in detail in \cite{Daverio:2016hqi}, in the presence of a {\it single} perfect fluid one does not actually need to solve the evolution equations \eqref{eq:evolE} and \eqref{eq:evolPi}, because $E$ and $P_i$ are fully determined by the gravitational fields through the constraint equations \eqref{eq:H} and \eqref{eq:Hi}. With the latter and \eqref{eq:SijPF}, we can then completely eliminate the perfect fluid variables in the evolution equations of gravity \eqref{eq:Kdot} and \eqref{eq:qdot}, which therefore become a closed system for $h_{ij}$ and $K_{ij}$ that can be evolved independently. The constraint equations then simply become a definition of the energy and momentum density that can be inferred from the gravitational field configurations, and these quantities automatically satisfy energy-momentum conservation \eqref{eq:evolE} and \eqref{eq:evolPi}. Note that this manipulation seems to be possible only in the context of perturbative cosmology where $E + p$ has a non-zero ``background" value, because it appears in the denominator in \eqref{eq:SijPF}. However, as shown in \cite{Daverio:2016hqi}, if the perfect fluid obeys the weak energy condition $p \geq - \ro$, then 
\beq
S^{ij} = \left\{ \begin{array}{ccc} p h^{ij} + \frac{P^i P^j}{E + p} & {\rm if} & E + p > 0 \\ 0 & {\rm if} & E + p = 0 \end{array} \right. \, ,
\eeq
so no problem arises in voids. In the presence of more than one fluid, or other types of matter content, this elimination still works, but only for one of the perfect fluids and the rest of matter must be given evolution equations to close the system.

Finally, note that one can also distinguish between the ``mass" and ``internal" (``temperature") contributions to the energy density 
\beq
\ro = \ro_M + \ro_T \, ,
\eeq
which allows for a more general equation of state $p \equiv p(\ro_M,\ro_T)$. This is especially interesting for the description of baryons, which have non-trivial thermodynamics $p = p(\ro_T)$, as opposed to dark matter which can simply be treated with $p = 0$ at large enough scales. Having traded one field $\ro$ for two new independent ones $\ro_M$ and $\ro_T$, we need an additional evolution equation to close the system and the natural choice is the conservation of mass, i.e.
\beq
\na_{\mu} J^{\mu} = 0 \, ,
\eeq
where 
\beq
J^{\mu} = \ro_M V^{\mu} \, ,
\eeq
is the mass $4$-current. Just like $T^{\mu\nu}$, one can also decompose that current in the $n$-frame, to get the canonical mass density 
\beq
M := -n_{\mu} J^{\mu} \equiv W \ro_M \, ,
\eeq
and $3$-current
\beq
h^i_{\mu} J^{\mu} = M v^i \, .
\eeq
The mass conservation equation then gives us the closing equation for the $E, P^i, M$ system
\beq \label{eq:evolM}
\( \pa_t - \Lie_{\be} \) M = - D_i \( N M \, \frac{P^i}{E + p} \) + N K M  \, . 
\eeq
Equations \eqref{eq:evolE}, \eqref{eq:evolPi} and \eqref{eq:evolM}, along with $p \equiv p(\ro_M,\ro_T)$, fully determine the perfect fluid dynamics. Having introduced one extra degree of freedom, however, these data can no longer be completely determined through the constraint equations \eqref{eq:H} and \eqref{eq:Hi} alone.

\section{Raw linearized equations of motion} \label{app:rawlin}

Here we will use the definition \eqref{eq:Udef}. A set of useful expressions to have at hand is
\bea
\hat{\ga}^{ab} & = & r_k^{-2} \[ q^{ab} \( 1 - \( 2 r_k^2 \ch + D^2 \bar{\ch} \) \) - 2 \( \bar{D}^{ab} \bar{\chi} + \ti{D}^{ab} \ti{\chi} \) \]   \, , \\
\hat{\Ga}^c_{ab} & = & \Ga^c_{ab}[q] + \[ 2 \de^c_{(a} D_{b)} - q_{ab} D^c \] \( r_k^2 \ch + \frac{1}{2}\, D^2 \bar{\ch} \)  \\
 & & + \[ 2 D_{(a} \bar{D}_{b)}^c - D^c \bar{D}_{ab} \] \bar{\ch} + \[ 2 D_{(a} \ti{D}_{b)}^c - D^c \ti{D}_{ab} \] \ti{\ch} \, , \nn \\
\hat{\cR} & = & 2 r_k^{-2} \[ 1 - \( 2 + D^2 \) r_k^2 \ch \]  \, , \label{eq:cRhatlin} \\
\hat{\cC}_{ab} & = & r_k^2 \[ q_{ab} \( \cX + r_k^2 \( \ch' + 4 \cX \ch \) + \frac{1}{2}\, D^2 \( \bar{U} + 2 \cX \bar{\ch} \) \) + \bar{D}_{ab} \( \bar{U} + 2 \cX \bar{\ch} \) + \ti{D}_{ab} \( \ti{U} + 2 \cX \ti{\ch} \) \] \, , \label{eq:cCabhatlin} \\
\hat{\cC}_a^b & = & \de_a^b \[ \cX + r_k^2 \( \ch' + 2 \cX \ch \) + \frac{1}{2}\, D^2 \bar{U} \] + \bar{D}_a^b \bar{U} + \ti{D}_a^b \ti{U}  \, , \\
\hat{\cC} & = & 2 \[ \cX + r_k^2 \( \ch' + 2 \cX \ch \) + \frac{1}{2}\, D^2 \bar{U} \]  \, , \label{eq:cChatlin} \\
\hat{\cK} & = & 2 r_k^2 \ka + D^2 \bar{\ka}   \, , \\
\hat{\cS} & = & 2 r_k^2 \pi + D^2 \bar{\pi}    \, , 
\eea
and we have used the fact that, on a scalar $\ph$
\beq
D^b \bar{D}_{ab} \ph \equiv \frac{1}{2}\, D_a \[ 2 + D^2 \] \ph \, , \hspace{1cm} D^b \ti{D}_{ab} \ph \equiv \frac{1}{2}\, \ti{D}_a \[ 2 + D^2 \] \ph \, .
\eeq
With the above equations, the background equations \eqref{eq:bgeq} and the background angular Laplacian definition \eqref{eq:D2rk2}, equations \eqref{eq:dothUp} to \eqref{eq:dotcPa} lead to the linear evolution equations
\bea
\dot{\ph} + \ph' & = & \te - \psi' - 2 \cX \psi + \cH \psi  \, , \label{eq:dotph} \\
\dot{\bar{u}} & = & \bar{\al} + \psi  \, , \label{eq:dotub} \\
\dot{\ti{u}} & = & \ti{\al}  \, , \label{eq:dotut} \\
\dot{\ch} + \ch' & = & \ka + \psi' + \cX \psi \, , \label{eq:dotch} \\
\dot{\bar{\ch}} & = & \bar{\ka} - \bar{U} \, , \label{eq:dotchb} \\
\dot{\ti{\chi}} & = & \ti{\ka} - \ti{U} \, , \label{eq:dotcht} \\
\nn\\
\dot{\te} + \te' + \cH \te & = & \psi'' + \( 4 \cX - \cH \) \psi' + \( \dot{\cH} - 2 \cH \cX + 2 \cX^2 - 2 k \) \psi + 2 \ph'' + 8 \cX \ph' + 4 \( \cX^2 - k \) \ph \nn \\
 & & +\, \ch'' + 5 \cX \ch' + \( 3 \cX^2 - 3 k - \frac{1}{2}\, \De^{(2)} \) \ch + \frac{1}{2}\, \De^{(2)} \( \bar{U}' + \cX \bar{U} \)  - \frac{1}{2} \( \si + \De^{(2)} \bar{\pi} \) - \pi  \, , \nn \\
  \label{eq:dotte} \\
\dot{\bar{\al}} + \bar{\al}' + 2 \cH \bar{\al} & = & 2 \( \cX^2 + k \) \bar{U} - 2 \( \psi + 2 \ph + \ch \)'  - 2 \cX \( \psi + 2 \ph + 2 \ch \) - 2 \bar{s} \, , \label{eq:dotalb} \\
\dot{\ti{\al}} + \ti{\al}' + 2 \cH \ti{\al} & = & \( 2 \cX^2 + 2 k + \De^{(2)} \) \ti{U} - 2 \ti{s} \, , \label{eq:dotalt} \\
\dot{\ka} + \ka' + 2 \cH \ka & = & -\, \psi'' - 3 \cX \psi' + 2 k \psi - 2 \ph'' - 6 \cX \ph' + 4 k \ph  \label{eq:dotka} \\
 & & -\, \ch'' - 4 \cX \ch' + \( 4 k + \De^{(2)} \) \ch - \De^{(2)} \( \bar{U}' + \cX \bar{U} \) + \pi \, , \nn \\
\dot{\bar{\ka}} + \bar{\ka}' + 2 \cH \bar{\ka} & = & \bar{U}' + 2 \cX \bar{U} + \psi + 2 \ph + \bar{\pi}  \, , \label{eq:dotkab} \\
\dot{\ti{\ka}} + \ti{\ka}' + 2 \cH \ti{\ka} & = & \ti{U}' + 2 \cX \ti{U} + \ti{\pi}  \, , \label{eq:dotkat}
\eea
the linear constraint equations 
\bea
3 \( \cH^2 + k \) \de & = & 6 \cH \te + 2 \cH \( 2 \ka + \De^{(2)} \bar{\ka} \) - 2 \[ \ph'' + 6 \cX \ph' + \( 6 \cX^2 - 2 k + \De^{(2)} \) \ph \]  \nn \\
 & & - \, 2 \[ \ch'' + 7 \cX \ch' + \( 9 \cX^2 - k + \frac{1}{2}\, \De^{(2)} \) \ch \] - \De^{(2)} \[ \bar{U}' + 3 \cX \bar{U} \]  \, , \label{eq:constrde} \\
2 \( - \dot{\cH} + \cH^2 + k \) v & = & 2 \( \te' + 2 \cX \te \) + 2 \( \ka' + 3\cX \ka \) + \De^{(2)} \( \bar{\ka}' + \cX \bar{\ka} \) - 2 \cH \( \ph' + 2\cX \ph \) + \frac{1}{2}\, \De^{(2)} \bar{\al} \, , \nn \\
\label{eq:constrv} \\
2 \( - \dot{\cH} + \cH^2 + k \) \bar{v} & = & 2 \te + \ka - 2 \cH \ph + \frac{1}{2} \( \bar{\al}' + 4 \cX \bar{\al} \) - \( \cX^2 + k \) \bar{\ka} \, , \label{eq:constrvb} \\
2 \( - \dot{\cH} + \cH^2 + k \) \ti{v} & = & \frac{1}{2} \( \ti{\al}' + 4 \cX \ti{\al} \) - \( \cX^2 + k + \frac{1}{2} \, \De^{(2)} \) \ti{\ka} \, , \label{eq:constrvt}
\eea
and the linear energy-momentum conservation equations
\bea
3 \( \cH^2 + k \) \( \dot{\de} + \de' + \cH \de \) & = & 2 \( \dot{\cH} - \cH^2 - k \) \[ v' + 4 \cX v +2 \ka + \De^{(2)} \( \bar{v} + \bar{\ka} \) \] \label{eq:dotde} \\
 & & -\, 6 \dot{\cH} \( \cH \de - \te - \cH \psi \) - \cH \( 3 \si + 2 \pi + \De^{(2)} \bar{\pi} \) \nn \\
 & & -\, 6 \( \cH^2 + k \) \( \psi' + 2 \cX \psi \) \, , \nn \\
2 \( - \dot{\cH} + \cH^2 + k \) \( \dot{v} + v' + 2 \cH v \) & = & 2 \( \ddot{\cH} - 2 \cH \dot{\cH} \) v - \si' - 2 \cX \si - \De^{(2)} \bar{s} \nn \\
 & & -\, 2 \( \dot{\cH} + 2 \cH^2 + 2 k \) \( \ph' + 2 \cX \ph \) \nn \\
 & & +\, 2 \( \dot{\cH} - \cH^2 - k \) \( \psi' + 2 \cX \psi \) + \cX \( 2 \pi + \De^{(2)} \bar{\pi} \) \, , \label{eq:dotv} \\
2 \( - \dot{\cH} + \cH^2 + k \) \( \dot{\bar{v}} + \bar{v}' + 2 \cH \bar{v} \) & = & 2 \( \ddot{\cH} - 2 \cH \dot{\cH} \) \bar{v} - \bar{s}' - 4 \cX \bar{s} - \( \cX^2 + k + \De^{(2)} \) \bar{\pi} - \pi - \si \nn \\
 & & +\, 2 \( \dot{\cH} - \cH^2 - k \) \psi - 2 \( \dot{\cH} + 2 \cH^2 + 2 k \) \ph \, , \label{eq:dotvb} \\
 2 \( - \dot{\cH} + \cH^2 + k \) \( \dot{\ti{v}} + \ti{v}' + 2 \cH \ti{v} \) & = & 2 \( \ddot{\cH} - 2 \cH \dot{\cH} \) \ti{v} - \ti{s}' - 4 \cX \ti{s} - \( \cX^2 + k + \frac{1}{2} \, \De^{(2)} \) \ti{\pi}  \, . \label{eq:dotvt}
\eea
Note that for \eqref{eq:dotte} we have used the Hamiltonian constraint \eqref{eq:EconstrC} to eliminate $E$ from \eqref{eq:dothTe}.

\bibliographystyle{JHEP}
\bibliography{mybib}

\end{document}

%% file: mydefs.tex
\newcommand{\beq}{\begin{equation}}
\newcommand{\eeq}{\end{equation}}
\newcommand{\bea}{\begin{eqnarray}}
\newcommand{\eea}{\end{eqnarray}}
\newcommand{\ben}{\begin{enumerate}}
\newcommand{\een}{\end{enumerate}}

\newcommand{\pa}{\partial}
\newcommand{\bo}{\square}
\newcommand{\na}{\nabla}
\newcommand{\ed}{{\rm d}}

\newcommand{\ti}{\tilde}
\newcommand{\Rs}{\mathbb{R}}

\newcommand{\para}{{\parallel}}
\newcommand{\ring}{\mathring}

\newcommand\rring[1]{%
  {
   \mathop{\kern0pt #1}\limits^{
     \vbox to-1.85ex{
       \kern-2ex 
       \hbox to 0pt{\hss\normalfont\kern.1em \r{}\kern-.45em \r{}\hss}%
       \vss 
     }
   }
  }
}

\renewcommand\({\left(}
\renewcommand\){\right)}
\renewcommand\[{\left[}
\renewcommand\]{\right]}
\newcommand{\bra}{\langle}
\newcommand{\ket}{\rangle}

\newcommand{\nn}{\nonumber}

\newcommand{\al}{\alpha}
\newcommand{\be}{\beta}
\newcommand{\ga}{\gamma}
\newcommand{\Ga}{\Gamma}
\newcommand{\de}{\delta}
\newcommand{\De}{\Delta}

\newcommand{\vep}{\varepsilon}

\newcommand{\te}{\theta}

\newcommand{\Te}{\Theta}

\newcommand{\ka}{\kappa}

\newcommand{\ro}{\rho}
\newcommand{\si}{\sigma}
\newcommand{\Si}{\Sigma}
\newcommand{\ta}{\tau}
\newcommand{\ph}{\phi}
\newcommand{\vph}{\varphi}
\newcommand{\ch}{\chi}
\newcommand{\om}{\omega}
\newcommand{\Om}{\Omega}
\newcommand{\Up}{\Upsilon}

\newcommand{\Lie}{{\cal L}}
\newcommand{\Ord}{{\cal O}}
\newcommand{\cH}{{\cal H}}

\newcommand{\cC}{{\cal C}}

\newcommand{\cK}{{\cal K}}
\newcommand{\cR}{{\cal R}}
\newcommand{\cP}{{\cal P}}
\newcommand{\cS}{{\cal S}}

\newcommand{\cM}{{\cal M}}

\newcommand{\cX}{{\cal X}}